%% file: main_manuscript.tex
\documentclass[twoside,11pt]{article}

\usepackage{blindtext}
\usepackage{amsmath}
\usepackage{subcaption}
\usepackage{multirow}
\usepackage{makecell}
\usepackage[dvipsnames]{xcolor}
\usepackage[preprint]{jmlr2e}

\input{makros}

\usepackage{lastpage}
\firstpageno{1}

\begin{document}

\newcounter{savecntr}

\title{Semi-supervised learning and integration of multi-sequence MR-images for carotid vessel wall and plaque segmentation}

\author{\name Marie-Christine Pali\textsuperscript{1,2,3} \email marie-christine.pali@i-med.ac.at \\
        \vspace{-.3cm}
        \AND
        \name Christina Schwaiger\textsuperscript{3} \email christina.schwaiger@vascage.at \\
        \vspace{-.3cm}
        \AND
        \name Malik Galijasevic\textsuperscript{1,2} \email malik.galijasevic@i-med.ac.at \\ 
        \vspace{-.3cm}
        \AND
        \name Valentin K. Ladenhauf\textsuperscript{1,2} \email valentin.ladenhauf@i-med.ac.at \\ 
        \vspace{-.3cm}
        \AND
        \name Stephanie Mangesius\textsuperscript{1,2} \email stephanie.mangesius@i-med.ac.at \\ 
        \vspace{-.3cm}
        \AND
        \name Elke R. Gizewski\textsuperscript{1,2} \email elke.gizewski@i-med.ac.at \\
        }

\institutions{
    \vspace{.1cm} \\
    \hspace*{.2cm} \textsuperscript{1} Department of Radiology, Medical University of Innsbruck. \\
    \hspace*{.2cm} \textsuperscript{2} Neuroimaging Research Core Facility, Medical University of Innsbruck. \\
    \hspace*{.2cm} \textsuperscript{3} Department of Data Science, VASCage - Centre on Clinical Stroke Research. 
}

\editor{My editor}

\maketitle

%% Abstract
\begin{abstract}
The analysis of carotid arteries, particularly plaques, in multi-sequence Magnetic Resonance Imaging (MRI) data is crucial for assessing the risk of atherosclerosis and ischemic stroke. In order to evaluate metrics and radiomic features, quantifying the state of atherosclerosis, accurate segmentation is important. However, the complex morphology of plaques and the scarcity of labeled data poses significant challenges. In this work, we address these problems and propose a semi-supervised deep learning-based approach designed to effectively integrate multi-sequence MRI data for the segmentation of carotid artery vessel wall and plaque.
The proposed algorithm consists of two networks: a coarse localization model identifies the region of interest guided by some prior knowledge on the position and number of carotid arteries, followed by a fine segmentation model for precise delineation of vessel walls and plaques. To effectively integrate complementary information across different MRI sequences, we investigate different fusion strategies and introduce a multi-level multi-sequence version of U-Net architecture. To address the challenges of limited labeled data and the complexity of carotid artery MRI, we propose a semi-supervised approach that enforces consistency under various input transformations.
Our approach is evaluated on 52 patients with arteriosclerosis, each with five MRI sequences. Comprehensive experiments demonstrate the effectiveness of our approach and emphasize the role of fusion point selection in U-Net-based architectures. To validate the accuracy of our results, we also include an expert-based assessment of model performance. Our findings highlight the potential of fusion strategies and semi-supervised learning for improving carotid artery segmentation in data-limited MRI applications.
\end{abstract}

\vskip 0.1in
%% Keywords

%\end{frontmatter}
\newpage

%% main text
%%
%%%%%%%%%%%%%%
\section{Introduction}\label{sec:intro}
%%%%%%%%%%%%%%

Stroke remains one of the most significant causes of death and disability worldwide, with carotid artery disease being a major contributing factor. Identifying atherosclerotic plaques within the carotid arteries that are prone to rupture and therefore pose a high risk of leading to stroke, is a critical aspect of stroke prevention and management. Magnetic resonance imaging (MRI) offers a non-invasive method of visualizing the carotid arteries, providing detailed imaging and allowing analysis of the vessel wall and plaque composition. In MRI, multiple sequences can be acquired, with each sequence capturing distinct tissue properties. Commonly used sequences for analyzing vessel wall and plaques within the carotid artery include T1-weighted (T1w), T1-weighted contrast-enhanced (T1ce), T2-weighted (T2w), proton density-weighted (PDw), and Time of Flight (TOF). Lipid-rich components typically appear mildly hyperintense on T1w images and isointense on T2w images, while calcifications are strongly hypointense across all sequences. The fibrous cap, aiding in the assessment of plaque stability, shows contrast enhancement on T1ce. TOF highlights flowing blood while suppressing stationary tissue signals, aiding in stenosis detection. Although this multi-modal data enriches the understanding of plaque characteristics, it also adds complexity to the segmentation tasks. Obtaining a reliable segmentation of these structures remains a critical challenge in medical imaging analysis, particularly when dealing with multiple sequences. 

Accurate segmentation of the carotid artery, its vessel wall, and associated plaques is essential to quantify plaque burden and assess the risk of stroke. Manual segmentation, while precise, is labor intensive and prone to inter-observer variability, especially when dealing with subtle features in the vessel wall and plaques. Automatic segmentation, on the other hand, poses significant challenges due to the variability in plaque morphology, imaging artifacts, and the limited availability of labeled data. Using deep learning-based methods for automated medical image segmentation has gained increased attention in recent years and has shown significant advances in various medical imaging applications, \citep{chwazh2022, masisr2023}. However, considering the complexity inherent in MR-image data, capturing pathological changes within vascular structures is still a challenging task, \citep{duzhhu2024, waya2023}. 
\\

In this work, we address the problem of segmenting carotid artery vessel walls and plaques in multi-sequence MRI data and propose a deep learning-based segmentation algorithm that leverages a semi-supervised learning approach to overcome the complexity and scarcity of (labeled) data. In particular, we propose a semi-supervised approach using consistency regularization that consists of two models: one for coarse localization of the carotid artery and another for fine segmentation of vessel walls and plaques. For the segmentation and localization modules we introduce a multi-level multi-sequence version of U-Net, incorporating an efficient fusion strategy and prior knowledge to localize the region of interest. By utilizing both labeled and unlabeled MRI sequences within a transformation consistent self-ensembling scheme, we aim to improve the model's generalization capabilities, making it more robust when faced with diverse plaque morphologies and varying image quality. To evaluate the performance of our approach, we conduct various experiments comparing different models and design choices, using five MRI sequences (T1w, T1ce, T2w, TOF, and PDw) from 52 patients.
%\\
The main contributions of this work are:

\begin{itemize}
    \item \textbf{Localization and fine segmentation:} We develop an automated method for the identification and segmentation of the carotid artery and pathological changes within the carotid artery in multi-sequence MRI data. For both, localization and fine segmentation, we introduce a multi-level multi-sequence version of U-Net architecture.
    
    \item \textbf{Semi-supervised segmentation for multi-sequence MRI data:} To address the scarcity and complexity of data, we propose a semi-supervised segmentation framework for multi-sequence MRI data. To the best of our knowledge, this is the first approach using semi-supervised learning for multi-sequence MRI carotid artery and atherosclerotic plaque segmentation.
    
    \item \textbf{Incorporation of prior knowledge:} To enhance localization accuracy, we introduce a novel prior-based approach with respect to the symmetry and number of connected components, guiding the identification of the region of interest (ROI).
    
    \item \textbf{Fusion strategy:} To effectively combine the information extracted from five different MRI sequences we analyze different fusion points for multi-sequence data and introduce an efficient embedding for U-Net-based architectures.
    
    \item The vessel wall and plaques can be segmented independently, allowing post-analysis to focus solely on the carotid artery plaque and assess its characteristics without requiring an additional identification network for separation.
\end{itemize}

The remainder of this paper is structured as follows. Section~\ref{sec:related_work} presents related work. The segmentation models and the semi-supervised learning approach are introduced in Section~\ref{sec:methods}, before the implementation details and experimental results are presented in Section~\ref{sec:experiments}. In Section~\ref{sec:discussion} we discuss different aspects of the proposed method and outline possible future directions. Finally, we conclude our work in Section~\ref{sec:conclusion}.

%%%%%%%%%%%%%%
\section{Related Work}\label{sec:related_work}
%%%%%%%%%%%%%%

Segmentation of medical images, particularly of vascular structures like the carotid arteries, has been an active area of research in recent years, driven by major advances in machine learning and deep learning. Traditional methods for medical image segmentation often rely on hand-crafted or statistical features, edge detection or regional similarities (e.g.~homogeneity in neighboring pixels), \citep{shag2010}. While these methods perform well on simpler tasks, they often struggle with the complexity and variability inherent in real-world medical data, \citep{khibhu2023}. With the advent of deep learning, convolutional neural networks (CNNs) have become the dominant approach for medical image segmentation, offering significant improvements in accuracy and robustness, \citep{kewara2017, xizhzo2024}. The U-Net architecture, first introduced in \citep{rofibr2015}, has seen widespread success in biomedical image segmentation due to its encoder-decoder structure, enabling both global and local feature extraction. Numerous adaptations have been proposed to improve the performance of U-Net, like residual connections, adaptations of convolutional blocks, different pooling and upsampling strategies or the integration of attention modules, \citep{zhli2017, khlo2020, oksc2018, wach2017}. 
\\
In the context of vascular segmentation, several works have applied deep learning-based techniques to identify and segment arteries or detect atherosclerotic plaques. For example, \citep{chch2023stroke} investigated pre-trained models such as YOLO \citep{redigi2015}, MobileNet \citep{hozhch2017}, and R-CNN \citep{gidoda2013}, to segment carotid plaque from MRI images that were already cropped to a region of interest (ROI) encompassing the carotid artery. Similarly, \citep{tssi2020} employed a U-Net architecture to segment carotid arteries from TOF images cropped to the region of interest. The 3D carotid arterial tree was then reconstructed using stacked 2D segmentations combined with morphological active contours. In \citep{wuxi2019} a deep U-shaped CNN with a weighted fusion layer was applied to segment both the lumen and the outer wall areas. The inputs of the U-Net were three ROIs of consecutive slices, interpreted as multi-channel input by the U-Net, to incorporate 2.5D information. The ROIs used within this approach were extracted via manually annotated center points. However, these models did not include a localization submodule and instead used manually cropped images. In contrast, \citep{chsu2020} implemented a YOLO detector for ROI localization. Following tracklet refinement and polar transformation of the cropped ROI, a CNN was used to address a boundary regression problem, segmenting the inner and outer walls of the carotid artery. \citep{wexi2022} used a centerline extraction algorithm on the MRI data to get a 2D ROI containing the carotid artery. The resulting ROIs were then passed to a U-Net, enhanced with residual blocks.
Similarly in \citep{weqi2022}, the vessel centerline was extracted to crop the images towards the carotid artery. The segmentation was based on two submodules, a U-Net and a network that considers the bottleneck layer and intermediate maps from the decoder to get a closer understanding on the edges of the segmentation. These approaches however, predominantly utilized a single MRI sequence, despite the fact that each sequence contains critical information regarding carotid artery plaque composition.

In \citep{zhwa2021}, a multi-sequence MRI approach was introduced, employing two 3D U-Net structures—one for whole MRI images and another for cropped patches—to obtain both local and global segmentation results. These results were integrated into a cascaded residual U-Net for final 3D segmentation. The authors in \citep{wayu2024} further advanced segmentation by using a 3D nn-Unet, \citep{isja2021}, to segment the carotid artery from multi-sequence MRI data. Their approach included localizing the bifurcation slice, cropping the images to a ROI, and concatenating the MRI sequences as a multi-channel input. Despite these advancements, in both works, \citep{zhwa2021} and \citep{wayu2024}, the information provided by all the different MRI sequences was utilized by concatenating them at the input-level of the CNN. Considering the findings for multi-modal brain tumor segmentation in \citep{ayhu2018}, this approach has however shown to be suboptimal. In particular, \citep{ayhu2018} evaluated various fusion strategies for multiple MRI sequences in CNNs and showed that late fusion was the most effective approach to maximize the information extracted from each sequence.

A limitation that was mentioned by the authors in the works above
is that their segmentation models heavily rely on large amounts of labeled data, which are often difficult to obtain in medical imaging applications. In particular, deep learning-based models that are trained in a supervised manner on a dataset that contains only a small number of labeled samples, tend to overfit. This means, that task-irrelevant features are considered for the classification or segmentation tasks, reducing the generalizability and robustness of the model. However, even the collection of unlabeled image data that can be used for training deep neural networks can face several problems like the rarity of some diseases or the protection of patient privacy. In addition, annotating medical imaging data is time-consuming, labor intensive and requires trained radiologists with several years of experience, \citep{hash2024}.
\\
To address the challenge of limited labeled data, semi-supervised learning has emerged as a promising solution \citep{hash2024}. By utilizing both labeled and unlabeled datasets, semi-supervised methods can improve the generalization and robustness of models while reducing the dependency on large amounts of annotated data. Techniques such as pseudo-labeling, consistency regularization, and entropy minimization have been successfully applied in medical imaging tasks to leverage unlabeled data more effectively, \citep{baaksi2017, yuwali2019, tava2017, liyuch2020, grbe2004}. Several studies have demonstrated the effectiveness of semi-supervised approaches for tasks like organ, lesion, or tumor segmentation, where only a small fraction of the dataset is labeled, \citep{feniwa2018, culili2019, raahna20223}. 
A comprehensive review of existing semi-supervised strategies for medical image segmentation is presented in \citep{hash2024}, including a comparative analysis of their impact on segmentation metrics relative to fully supervised approaches. The findings reveal that all semi-supervised paradigms enhance segmentation performance when trained with the same number of labeled samples, underscoring the efficacy of deep semi-supervised learning techniques in this domain.

%%%%%%%%%%%%%%
\section{Methods}\label{sec:methods}
%%%%%%%%%%%%%%

Considering all the findings and limitations mentioned in Section~\ref{sec:related_work}, in this work we aim to build upon these advances and introduce a semi-supervised learning approach for multi-sequence MRI carotid artery vessel wall and plaque segmentation.
In this section, we formally introduce the used notation as well as the considered image segmentation problem and present our proposed multi-level multi-sequence semi-supervised segmentation approach.

\paragraph*{Notation}

\begin{figure}[t]
    \centering
    \begin{subfigure}{0.3\linewidth}
    \includegraphics[width=\linewidth]{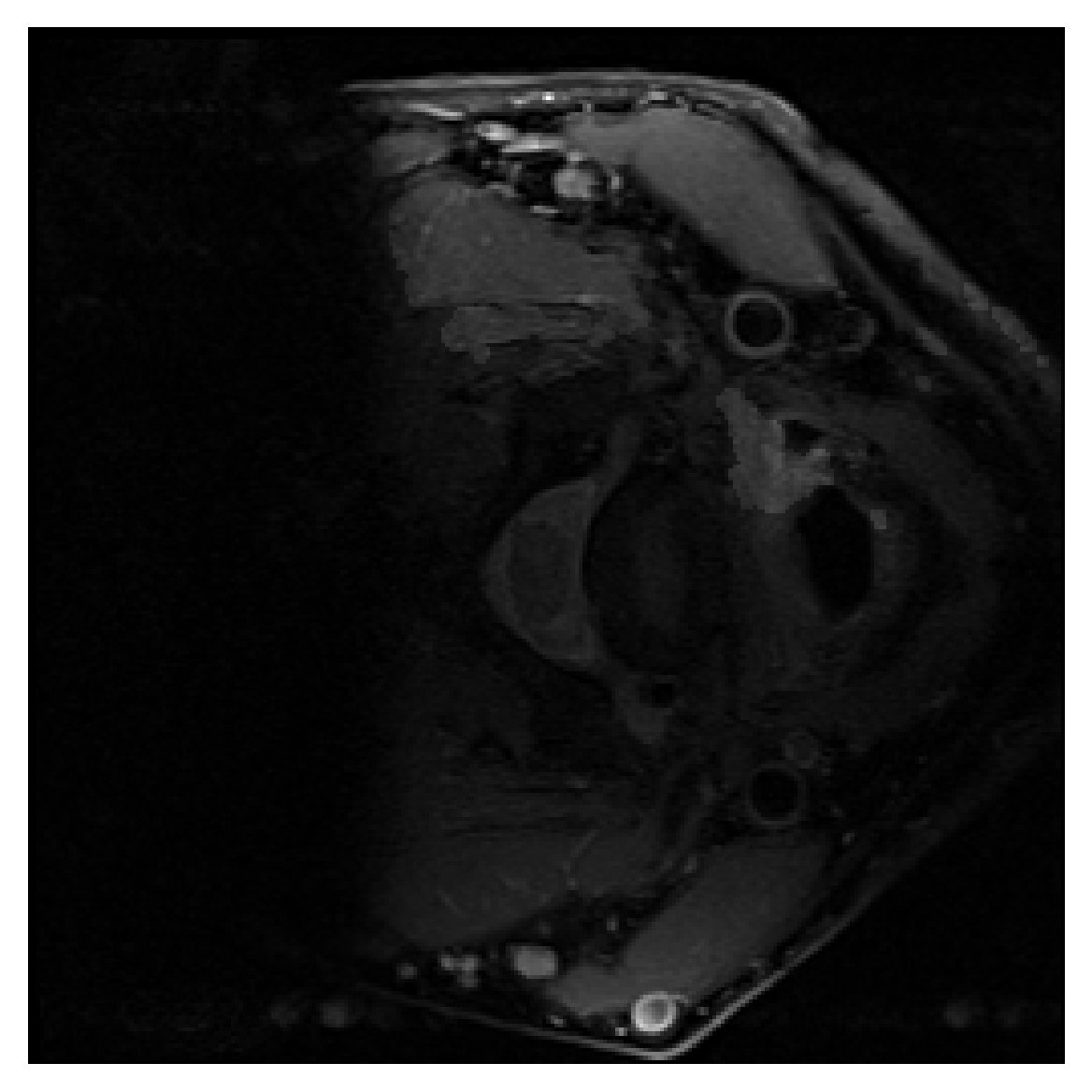}
    \caption{$x_{ik}^1$ - PDw}
    \end{subfigure}
    \begin{subfigure}{0.3\linewidth}
    \includegraphics[width=\linewidth]{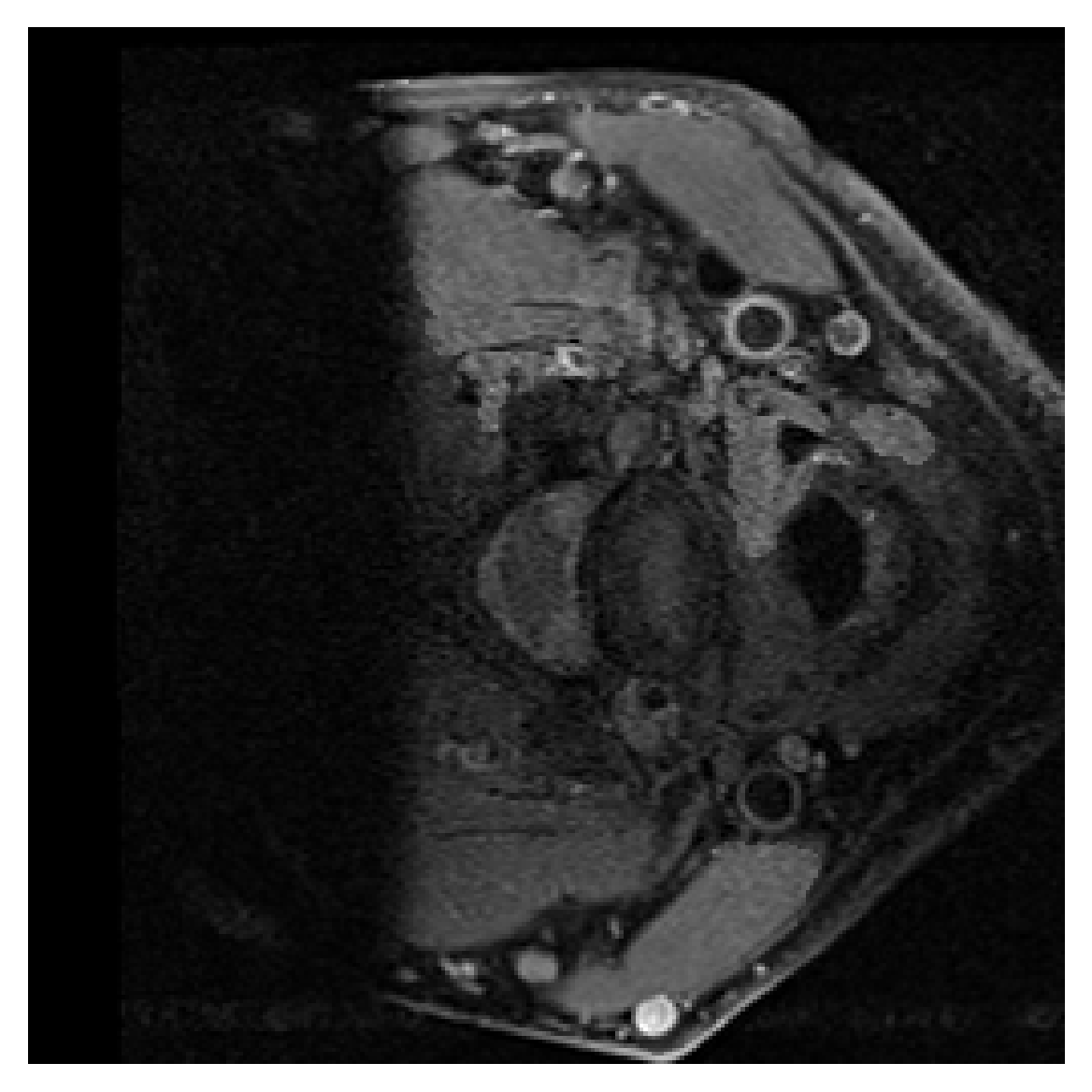}
    \caption{$x_{ik}^2$ - T1w}
    \end{subfigure}
    \begin{subfigure}{0.3\linewidth}
    \includegraphics[width=\linewidth]{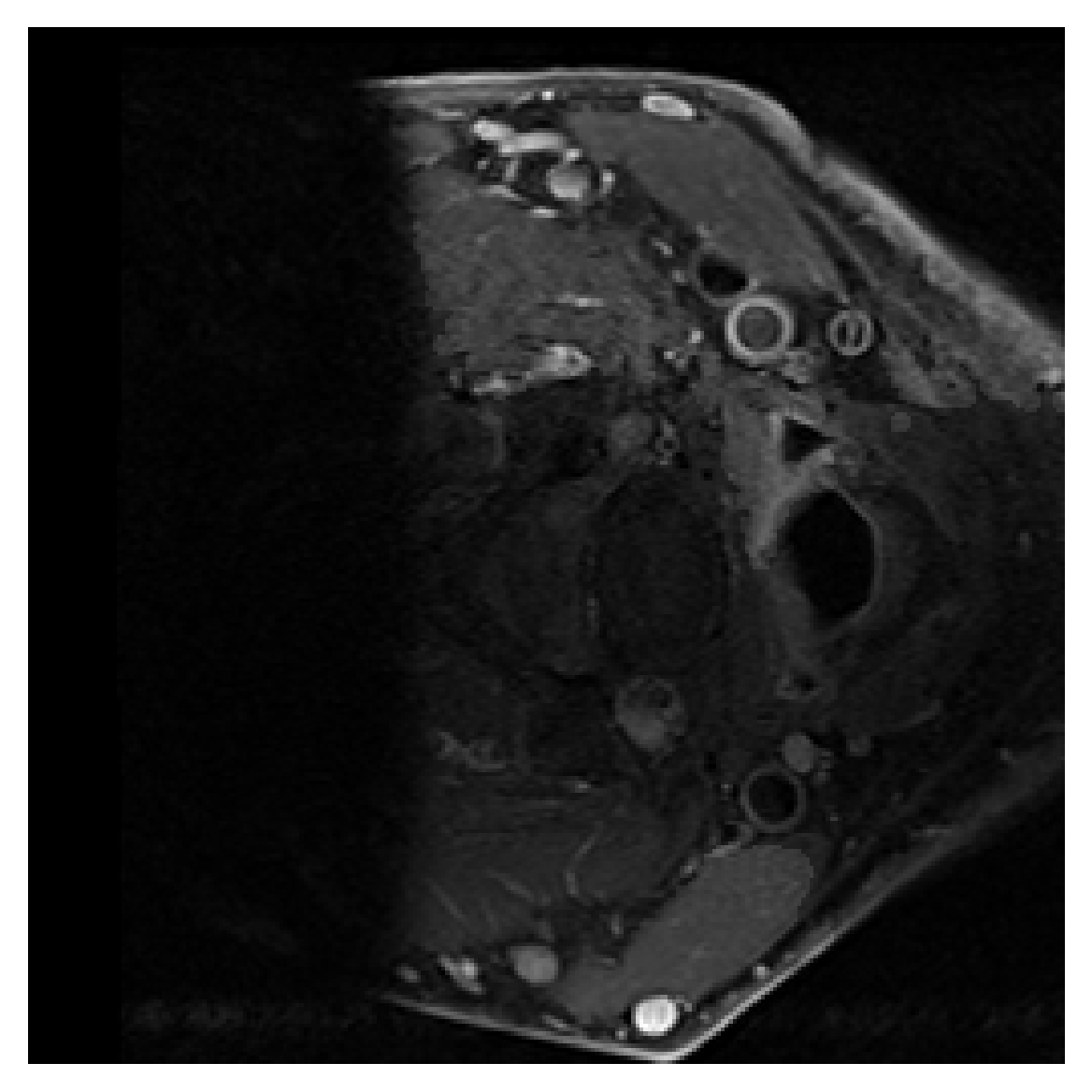}
    \caption{$x_{ik}^3$ - T1ce}
    \end{subfigure}
    \begin{subfigure}{0.3\linewidth}
    \includegraphics[width=\linewidth]{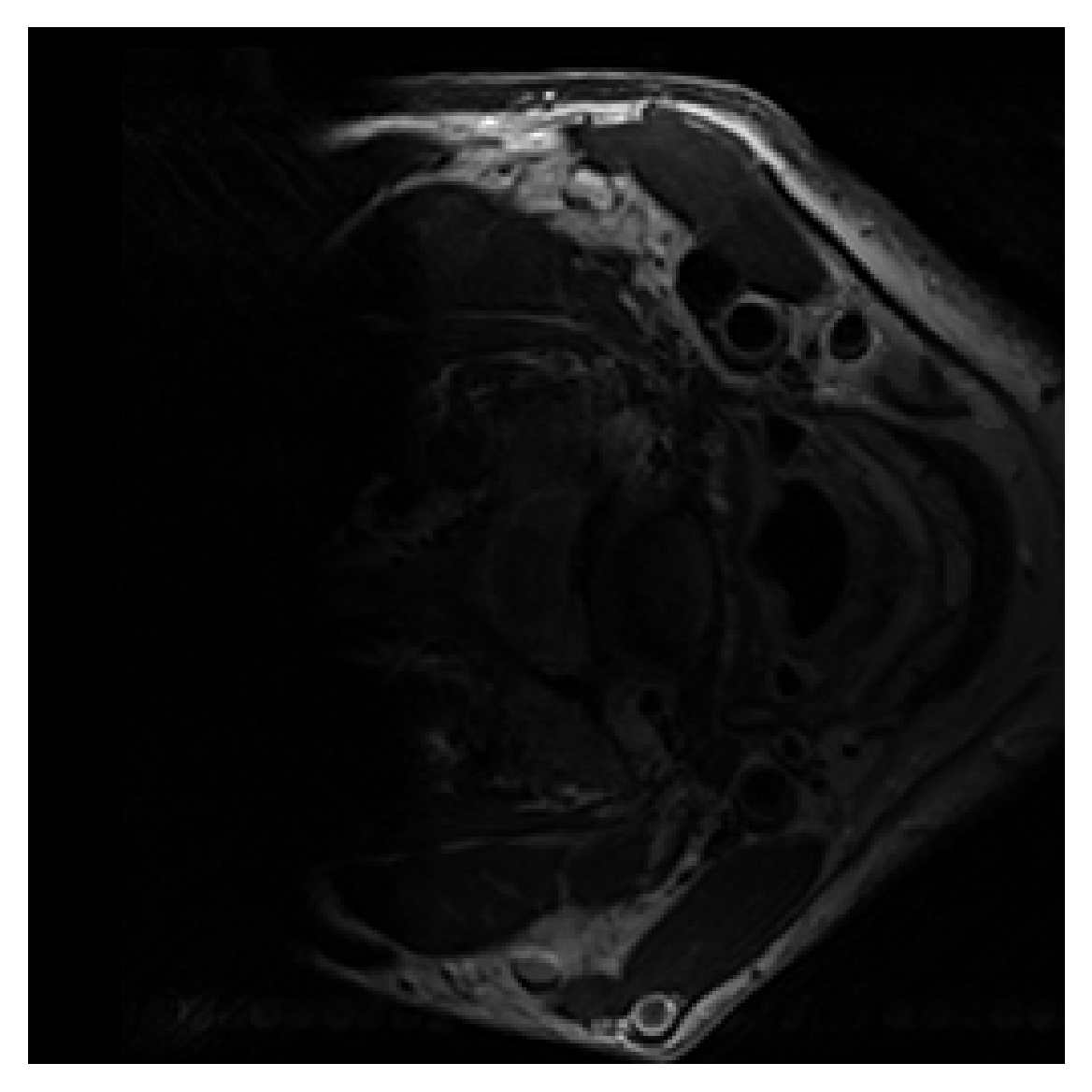}
    \caption{$x_{ik}^4$ - T2w}
    \end{subfigure}
    \begin{subfigure}{0.3\linewidth}
    \includegraphics[width=\linewidth]{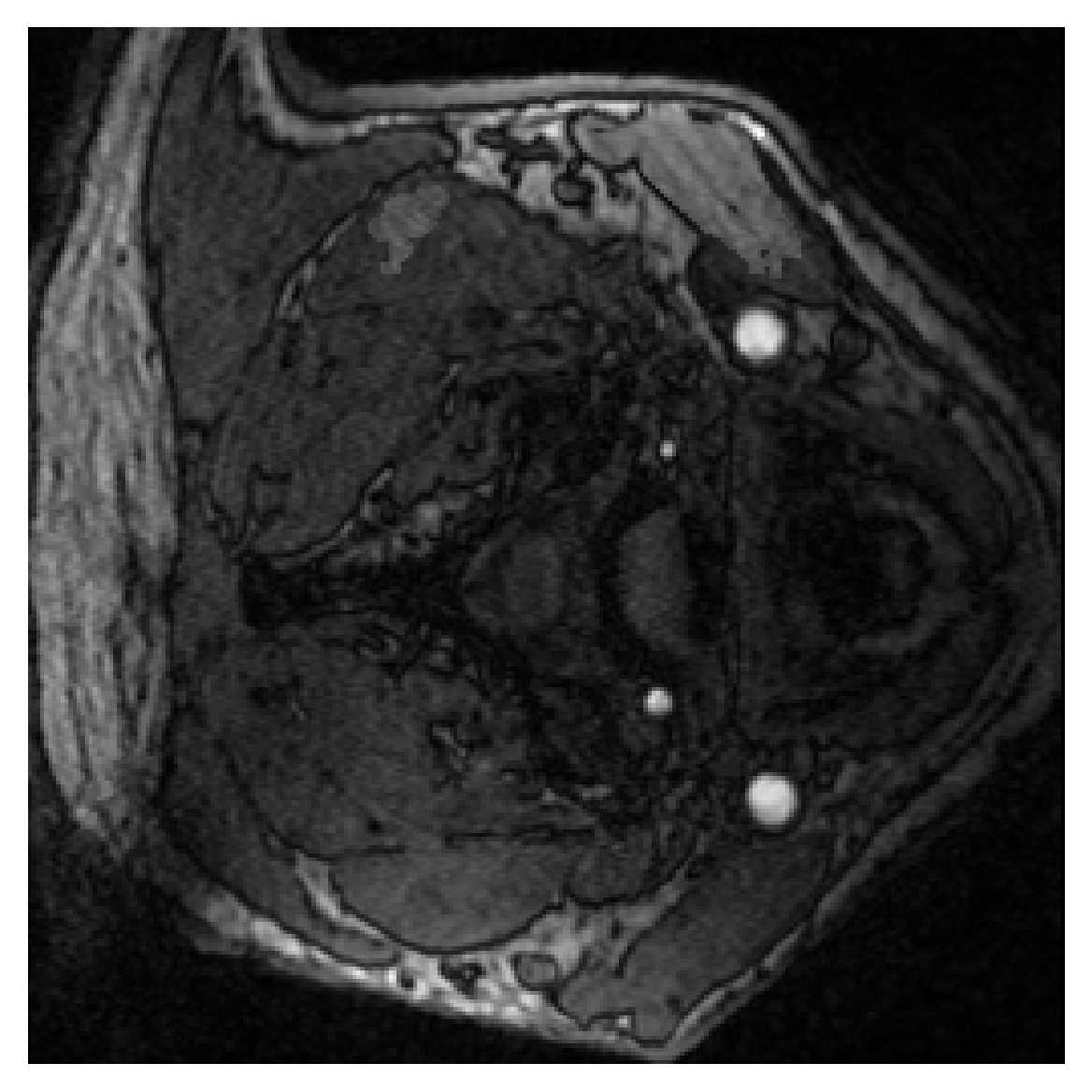}
    \caption{$x_{ik}^5$ - TOF}
    \end{subfigure}
    \caption{Slices $\{x_{ik}^j\}_{j=1}^5$ for a fixed patient $i$ and a slice index $k$ for the $5$ MRI sequences.}
    \label{fig:sequeces}
\end{figure}

We consider a clinical dataset of $n_p$ patients, where we refer to the set of patients as $N=\{1, \ldots, n_p\}$. For each patient, five different MRI sequences were collected that describe the 3D volume of the neck, where each sequence consists of multiple 2D image slices $\img\in\R^{\dimgh\times\dimgw}$ of size $\dimgh\times\dimgw$. Therefore, for each patient $i \in N$, we have $\{x_{ik}^1\}_{k \in K_i}$, $\{x_{ik}^2\}_{k \in K_i}$, $\{x_{ik}^3\}_{k \in K_i}$, $\{x_{ik}^4\}_{k \in K_i}$, $\{x_{ik}^5\}_{k \in K_i}$, where $K_i$ %is the set of all slices 
denotes the set of slice indices of patient $i$. We also refer to the slices of each sequence for a fixed patient $i$ as $X_i = \{(x_{ik}^1, x_{ik}^2, x_{ik}^3, x_{ik}^4, x_{ik}^5)\}_{k \in K_i}$. An example of the different sequences for a fixed patient $i$ and a fixed slice index $k \in K_i$ is depicted in Figure~\ref{fig:sequeces}. For some of the patients, the first sequence was partially labeled. This means, for a patient $i \in N_L \subseteq N$, where $N_L$ denotes the subset of patients for which some slices are annotated, there exist slice indices $\widetilde{K}_i \subseteq K_i$ for which segmentation masks $\{y_{k}\}_{k \in \widetilde{K}_i}$, with $y_{ik}\in\{ 0,1\}^{\dimgh\times\dimgw\times C}$ and $C$ the number of segmentation classes, are provided by experts from radiology.
The manual annotation was performed on one of the MRI sequences, which we fix to be the first one ($j=1$) without loss of generality. For this, $\{y_k\}_{k \in \widetilde{K}_i}$ denote the segmentation masks corresponding to $\{x_{ik}^1\}_{k \in \widetilde{K}_i}$. We denote the set of labeled slices as $X^L = \{(x_{ik}^1, x_{ik}^2, x_{ik}^3, x_{ik}^4, x_{ik}^5)\}_{k \in \widetilde{K_i}, i \in N}$, the set of unlabeled slices as $X^U = \{(x_{ik}^1, x_{ik}^2, x_{ik}^3, x_{ik}^4, x_{ik}^5)\}_{k \notin \widetilde{K_i}, i \in N}$, and by $X = \{(x_{ik}^1, x_{ik}^2, x_{ik}^3, x_{ik}^4, x_{ik}^5)\}_{k \in K_i, i \in N} = X^L \cup X^U$ the set of all labeled and unlabeled slices. Further, we refer to $X^j = \{x_{ik}^j\}_{k \in K_i, i \in N}$ as the set of all slices and patients of sequence $j$, and $X_i^j$ as the set of all slices of patient $i$ belonging to sequence $j$. 
By $Y = \{y_{ik}\}_{k \in \widetilde{K_i}, i \in N}$ we denote the set of all segmentation masks and by $Y_i$ the annotated slices of a patient $i$.
Additionally, given a statement $P$, let $\chi(\cdot)$ denote the indicator function such that $\chi(P) = 1$ if $P$ is true and $0$ otherwise, and $\mathbf{1}_{\dimgh\times\dimgw}$ a matrix of size $\dimgh\times\dimgw$ with every entry equal to one.
\\

Segmenting vessel wall and plaques within the carotid artery in multi-sequence MRI data poses significant challenges due to the morphological variability of plaques, imaging artifacts, and the limited availability of labeled data. Plaques exhibit diverse spatial and structural characteristics, and with only a small amount of labeled data available, supervised approaches struggle to generalize effectively. In order to address these challenges, we propose a semi-supervised learning framework that leverages both labeled and unlabeled data by using consistency regularization. This approach enforces predictive consistency under various input transformations, enhancing the model's robustness and generalization. In particular, achieving geometric consistency is non-trivial, as CNNs lack inherent equivariance to spatial transformations such as rotations. To overcome this limitation, we incorporate a transformation-consistent regularization scheme on the input and output space of our model similar to \citep{liyuch2020}, extending their results to multi-sequence data combined with a perturbed student one-way consistency approach as suggested by \citep{grorse2023}, and estimations on the uncertainty of predictions. 
By enforcing consistency across spatial transformations, we aim to strengthen the models ability to adapt to diverse morphology and positional variations of pathological changes within the carotid artery and further, utilize both labeled and unlabeled data more efficiently.
Given that we are working with multi-sequence data, we further have to address the challenge of combining all the information contained within each sequence. For this, we analyze different fusion points, allowing the model to effectively leverage complementary information provided by different modalities, and introduce a novel multi-level multi-sequence version of U-Net architecture.

\begin{figure*}[t]
    \centering
    \includegraphics[width=0.9\linewidth]{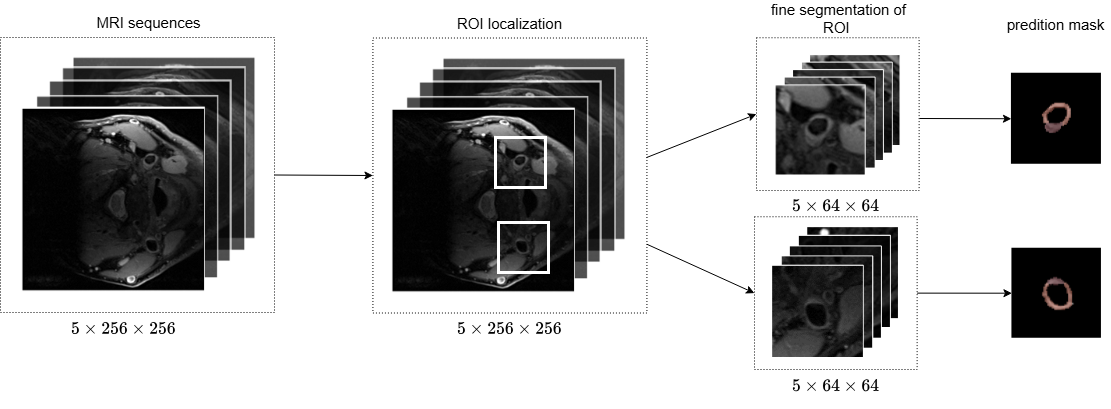}
    \caption{Segmentation workflow.}
    \label{fig:workflow}
\end{figure*}

Considering the vast amount of information contained in multi-sequence MRI data and the class imbalance present in each image slice, we formulate our problem as hierarchical segmentation problem. This means, our segmentation approach consists of two networks, for coarse localization and fine segmentation of vessel wall and plaque, respectively (see Figure~\ref{fig:workflow} for the overall workflow). The localization network is used to identify the region of interest and guided by some prior knowledge, followed by the fine segmentation to accurately delineate vessel walls and plaques within this localized area.

In the following, we will describe all these concepts and our proposed method in more detail.

\subsection{Model Architecture}

\begin{figure*}[ht]
    \centering
    \includegraphics[width=.9\linewidth]{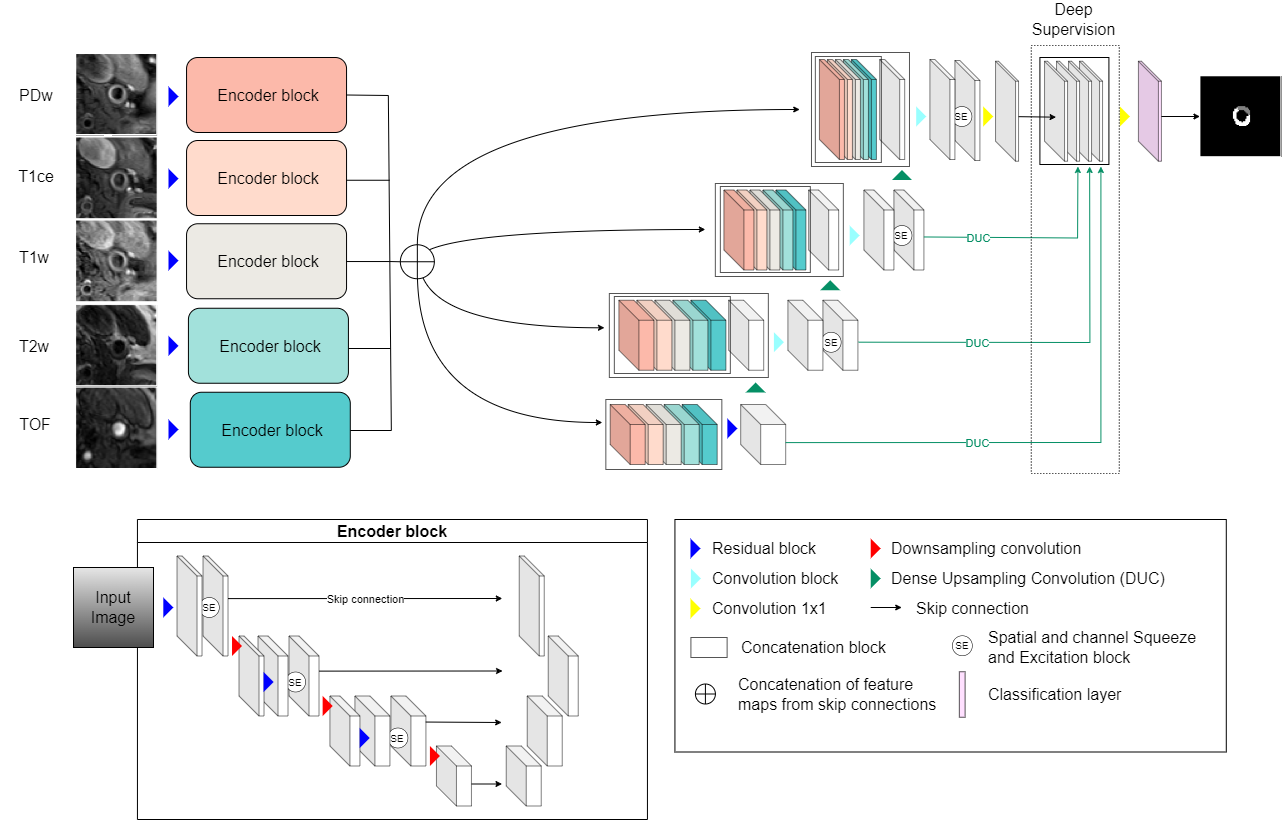}
    \caption{Model architecture of fine segmentation model.}
    \label{fig:fine-seg}
\end{figure*}

\noindent
For both networks, localization and fine segmentation, we propose a multi-level-multi-sequence version of U-Net. As opposed to standard U-Net, we replaced classical convolutional layers with residual blocks in the encoder path, in order to address the vanishing gradient problem and allow the network to better propagate information between low and high levels. Each residual block consists of two convolutional layers with instance normalization and Parametric ReLU (PReLU) activation, followed by a skip connection that adds the input of the block to its output. Instead of the traditional max-pooling operation, we use convolutional layers with a stride of $2$ for downsampling. Convolutional pooling has the advantage of being trainable, enabling the network to adapt the downsampling process based on the data.
In the decoder path we use dense upsampling convolution (DUC), \citep{wach2017}, instead of traditional bilinear interpolation or transposed convolutions to upsample feature maps and recover the spatial resolution. These layers help to maintain high-resolution features and ensure the preservation of fine details during the upsampling process. 
\\
To further improve the network's ability to focus on relevant features, we integrated Squeeze and Excitation (SE) blocks, \citep{hushsu2017}, within both encoder and decoder path of the fine segmentation model. These modules recalibrate the feature maps by adaptively weighting the importance of each channel and spatial location, enabling the model to selectively emphasize features that are most relevant for the segmentation task.
In addition to this, to guide the network during training and improve convergence, deep supervision is applied at multiple levels of the decoder. Specifically, segmentation maps are computed from the output of each decoder layer. These intermediate segmentation maps are then upsampled to the size of the final output using dense upsampling convolution. The upsampled segmentation maps are concatenated and further refined using a 2D convolutional layer, producing the final segmentation output. 
An overview of the fine segmentation model is depicted in Figure~\ref{fig:fine-seg}. For the coarse localization, we use a similar architecture but without SE-modules and deep supervision, focusing on computational efficiency and maintaining a simpler structure suited for identifying the region of interest with lower complexity.
\\\\
Given that we are working with multiple sequences for each patient, it is important to effectively combine the information from these different modalities. For that, we analyzed different fusion strategies within our proposed model, i.e.~fusion at input level and fusion at bottleneck. Although most published works, dealing with multiple MR-image sequences, use early stage fusion, usually at the input level, we observed that fusing image-sequence feature maps at later stages yielded better results. This observation seems to be consistent with the results presented in \citep{ayhu2018}, which analyzed different fusion strategies for non-U-Net-shaped CNNs, and is probably due to the fact that more high level information can be preserved when each sequence is processed separately. For this, in our model, we decided to fuse all the information in the final layer. This means, each input sequence is processed through its own independent encoder path, allowing the network to learn distinct features from each modality before fusing the information in the bottleneck. In particular, this process combines the rich feature representations from all MRI sequences, allowing the model to leverage the complementary information provided by different modalities.
Given that our model is based on a U-Net structure, at each stage in the decoder, skip connections transfer feature maps from the corresponding encoder layers. Specifically, these skip connections concatenate the feature maps from the encoder stages with the decoder stages, allowing the network to use both high-level and low-level features for precise segmentation. This concatenation aids the network in recovering fine details that may have been lost during downsampling processes. In order to integrate this concept in our multi-sequence setting and effectively use the information provided by all sequences, we concatenate the feature maps of each sequence and forward the concatenated map to the decoder. In particular, each sequence is processed by a separate encoder block and the resulting feature maps of each level are forwarded to the next level in the encoder and to the same level in the decoder. This means that $m$ feature maps are sent to each decoder level, where $m$ denotes the number of sequences. In each decoder level the $m$ feature maps are concatenated with the upsampled decoder map from the lower level. Overall, if the bottleneck is the fusion point, the information is not only concatenated in the bottom level (bottleneck), but in each decoder level (see Figure~\ref{fig:fine-seg}). This possibly also explains the superior behavior with respect to other fusion strategies.

\subsection{Semi-supervised segmentation}

In order to address the challenges associated with the variability in plaque morphology, imaging artifacts and also the limited availability of labeled data, for our semi-supervised approach we use consistency regularization within a Student-Teacher framework. In particular, this approach consists of two models which are enforced to make consistent predictions under different transformations of the input images, enabling not only the utilization of unlabeled data but also enhancing the model's ability to generalize across diverse anatomical and pathological presentations. Considering the findings in \citep{grorse2023}, we follow a one-way consistency approach where only the input images for the student model are perturbed, while the teacher model processes the input images without alterations. Perturbations that are usually applied in this context include photometric transformations like noise or intensity transformations and network dropout. The reason for this is that convolutions are invariant to this kind of perturbations, however not equivariant to transformations that affect the spatial layout of the input images. This means, for some CNN model $\modelS$ and for every geometric transformation 
$T_\gamma^\mathrm{G}$ of the input image $x_{ik}^j$, for some patient $i$, slice $k$ and sequence $j$, there exists another geometric transformation $T_{\gamma'}^\mathrm{G}$ of the output, such that $\modelS(T_\gamma^\mathrm{G}(x_{ik}^j)) = T_{\gamma'}^\mathrm{G}(\modelS(x_{ik}^j))$, however in general $T_\gamma^\mathrm{G} \neq T_{\gamma'}^\mathrm{G}$.

This limits the application of stronger regularization using a wider range of data transformations in segmentation tasks. In order to enhance regularization using consistency, we therefore incorporated a transformation-consistent regularization scheme on the input and output space of our model similar to \citep{liyuch2020}, to take advantage of both photometric and geometric perturbations. By enforcing consistency across spatial transformations, such as rotations and translations, alongside standard photometric alterations, we encourage the model to maintain stable predictions under a wider range of input conditions. This approach not only addresses the limitations of purely photometric transformations but also strengthens the models robustness to variations in spatial structure, ultimately enhancing its generalization capabilities.

In addition to this, in order to ensure that the student model learns only from meaningful and reliable predictions of the teacher model when calculating the consistency loss, we further incorporated the idea of an uncertainty aware scheme similar to \citep{yuwali2019}. 
Besides generating the predictions for the input data, the teacher model also estimates the uncertainty for the predictions using Monte Carlo sampling. The student model is hence optimized by the consistency loss using only confident predictions from the teacher model guided by the estimated uncertainty. 
\\

\paragraph{Semi-supervised segmentation}
To define our semi-supervised segmentation approach, let $\Tgp = \Tgpdef$ denote a randomly selected composition of various geometric $\Tg$ and photometric $\Tp$ perturbations consisting of strong and weak scaling, rotation, random crop, random perspective, noise, sharpness and intensity transformations as well as random network dropout in the encoder path. The input images for the student model are perturbed using $\Tgp$, defined by the perturbation parameters $\paramGP = (\paramG, \paramP)$. For the given geometric perturbation parameter $\paramG$, defining all the applied geometric transformations of $\Tgp$, the transformation $\Tg$ is also applied to the output of the teacher model when calculating the consistency loss. The transformation consistency is obtained by applying geometric transformations to the input of the student model and also the output of the teacher model. The model hence learns how to approximate these perturbations at input and output space. 
\\

\begin{figure}[ht]
    \centering
    \includegraphics[width=1.0\linewidth]{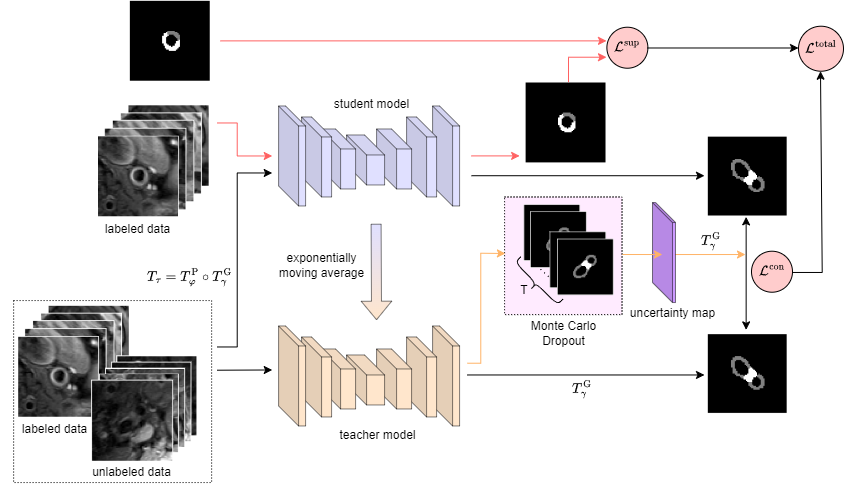}
    \caption{One-way consistency semi-supervised learning approach with perturbed student and clean teacher.}
    \label{fig:semi-sup}
\end{figure}

\noindent
The training of our model can thus be formulated as minimizing the following objective function with respect to the network parameters $\paramS$ of the student model,
\begin{align*}
    \Ltotal_t(\modelS(\Xlu), \Y) = \Lsup_t(\modelS(\Xl), \Y) +\lambda(t)\Lcon_t(\modelS(\Tgp(\Xlu)), \Tg(\modelT(\Xlu))) , 
\end{align*}
where $\Lsup$ denotes the supervised loss, $\Lcon$ the unsupervised consistency loss and $\paramT$ the network parameters of the teacher model. $\lambda(t)$ denotes a weighting ramp-up function depending on the training step $t$, regulating the contribution of the consistency term. An overview of our proposed framework is depicted in Figure~\ref{fig:semi-sup}.
\\\\
In the considered consistency objective, gradients are only calculated for the student model. The teacher model is defined to be an ensemble of consecutive student models. For this, the parameters $\paramTt$ of the teacher model at training step $t$ are defined as the exponential moving average (EMA) of successive $\paramS$ weights,
\begin{align*}
    \paramTt = \alpha\paramTtm + (1-\alpha)\paramSt ,
\end{align*}
where $\alpha$ denotes a momentum term that controls how far the ensemble reaches into training history, \citep{tava2017}. Using this temporal ensemblig of student weights has shown to stabilize training trajectories and improve prediction results. Since training in consistency based semi-supervised learning can be less stable, this EMA-teacher framework can help with its improved generalizability, \citep{atfiiz2019, izpoga2018}.
The weighting function $\lambda(t)$ is defined as 
\begin{align}\label{equ:rampup}
    \lambda(t) = k \exp\left(-5 \left(1-\frac{\min \{ t, R\}}{R}\right)^2 \right) ,
\end{align}
where $t$ denotes the current training step and $R$ the ramp-up length, i.e. the training step where $\lambda(t) = k$. This function allows a warm start, ensuring that the contribution of the unsupervised consistency loss term is slowly increased and the supervised loss dominates the beginning of the training. 
\\\\
The supervised loss is calculated on labeled data $\Xl$ whereas the unsupervised loss is calculated using both labeled and unlabeled data $\Xlu = \Xl \cup X^\text{U}$. For the supervised loss, we use two different loss functions for coarse localization and fine segmentation, respectively. The coarse localization is defined as binary classification problem, classifying pixels inside and outside the carotid artery as foreground and background, respectively. For that, within the coarse localization setting, we use a weighted combination of a modified Tversky loss and binary cross entropy loss,
\begin{align*}
    \Lsup(\pred, \gt) = \lambda_\mathrm{loc}\mathcal{L}_\mathrm{mT}(\pred, \gt)  + (1-\lambda_\mathrm{loc})\mathcal{L}_\mathrm{BCE}(\pred, \gt) ,
\end{align*}
where $\pred$ denotes the probabilities obtained from the network, $\gt$ the corresponding ground truth mask and $\lambda_\mathrm{loc}\in \left[ 0,1\right]$ a weighting parameter controlling the relative contribution of both losses. The modified Tversky loss is defined as $\mathcal{L}_\mathrm{mT}(\pred, \gt) = 1 - \mathrm{mTI}_{\delta}(\pred, \gt)$, with
\begin{align}\label{equ:mTI}
    \mathrm{mTI}_\delta(\pred, \gt) =  \frac{\sum_{i=1}^M \gt_i \pred_i}{\sum_{i=1}^M \gt_i \pred_i + \delta\sum_{i=1}^M \gt_i (1-\pred_i) + (1-\delta)\sum_{i=1}^M (1-\gt_i) \pred_i} ,
\end{align}
where $\gt_i$ denotes the ground truth value for pixel $i$, $\pred_i$ the predicted probability for pixel $i$, $M = \dimgh\cdot\dimgw$ the number of all pixels within the predicted and ground truth mask, and $\delta\in \left[ 0,1\right]$ a weighting coefficient for false positive and false negative examples. 
The binary cross entropy loss is defined as 
\begin{align*}
    \mathcal{L}_\mathrm{BCE}(\pred, \gt) = -\frac{1}{M} \sum_{i=1}^M\left[ \gt_i \log(\pred_i) + (1-\gt_i)\log(1-\pred_i)\right] .
\end{align*}
For the fine segmentation we consider a multiclass segmentation problem (vessel wall, plaque and background). In order to focus on harder-to-classify regions while accounting for the class imbalance, we decided to use the Asymmetric Unified Focal loss introduced in \citep{yesasc2022}, which combines a modified Focal and Focal Tversky loss with asymmetry parameters to prioritize underrepresented classes. For the supervised loss within the fine segmentation, we thus use
\begin{align*}
    \Lsup(\pred, \gt) = \lambda_\mathrm{seg}\mathcal{L}_\mathrm{maF}(\pred, \gt)  + (1-\lambda_\mathrm{seg})\mathcal{L}_\mathrm{maFT}(\pred, \gt) ,
\end{align*}
where $\mathcal{L}_\mathrm{maF}$ denotes the modified asymmetric Focal loss, $\mathcal{L}_\mathrm{maFT}$ the modified asymmetric Focal Tversky loss and $\lambda_\mathrm{seg}$ a weighting parameter of the two loss functions. 
Let $r$ denote the rare classes (vessel wall and plaque) and $b$ the background class. The modified asymmetric Focal loss is defined as
\begin{align*}
    \mathcal{L}_\mathrm{maF}(\pred, \gt) = -\frac{\delta}{M}\sum_{i=1}^M\sum_{c=r}\gt_i^c \log(\pred_i^c) - \frac{1-\delta}{M}\sum_{i=1}^M (1-\pred_i^{b})^\vartheta \gt_i^{b} \log(\pred_i^{b}) ,
\end{align*}
and the modified asymmetric Focal Tversky loss as
\begin{align*}
    \mathcal{L}_\mathrm{maFT}(\pred, \gt) = \sum_{c\neq r}(1-\mathrm{mTI}_\delta^c(\pred, \gt)) + \sum_{c=r}(1-\mathrm{mTI}_\delta^c(\pred, \gt))^{1-\vartheta} ,
\end{align*}
where $\mathrm{mTI}_\delta^c$ denotes the modified Tversky Index ($\mathrm{mTI}_\delta$) in (\ref{equ:mTI}) for class $c$, delta again the weighting parameter controlling the relative contribution of positive and negative examples and $\vartheta$ a hyperparameter controlling both suppression of the background class and enhancement of the rare class.
\\\\
In order to incorporate also unlabeled data, we make use of a transformation consistent regularization scheme. To enforce transformation consistency, we aim to minimize the difference between the transformed teacher network output and the student model predictions of the transformed inputs using a mean squared error (MSE) loss. Let $v^\gamma\in \{ 0,1\}^{\dimgh\times\dimgw\times C}$ denote a validity mask, defined by $v^\gamma =\chi(\Tg(\mathbf{1}_{\dimgh\times \dimgw\times C}) = 1)$, ensuring that the calculation of this difference is unaffected by padding after applying some geometric transform $\Tg$. The consistency loss is defined as
\begin{align}\label{equ:consistency}
    \Lcon(\modelS(\Tgp(\Xlu)), \Tg(\modelT(\Xlu))) = \frac{1}{\sum v^\gamma} \lVert v^\gamma (\modelS(\Tgp(\Xlu)) -\Tg(\modelT(\Xlu))) \rVert_2^2 , 
\end{align}
where $\Xlu$ includes both labeled and unlabeled samples. Note that by using only photometric transformations and network dropout, this loss reduces to the simple MSE loss between the predictions on perturbed inputs from the student model $\modelS(\Tp(\Xlu))$ and the predictions on clean inputs from the teacher model $\modelT(\Xlu)$, using exponentially moving average weights of the student model. By additionally including geometric transformations $\Tg$, affecting the spatial layout of the input images, these transformations have also to be included in the loss calculation. Given that CNNs are in general not equivariant to geometric transformations, the model has to be regularized to be transformation consistent under this kind of perturbations. Using the consistency loss term defined in \eqref{equ:consistency} this is achieved by incorporating a transformation consistent regularization on input and output space of the model. In particular, using \eqref{equ:consistency} the model learns how to approximate the geometric transformations applied to the outputs of the teacher model to the geometric transformations applied to the inputs of the student model. The model thus learns transformation equivariance during training while additionally enhancing its robustness and generalization capabilities using both labeled and unlabeled data.
\\

\paragraph{Uncertainty estimation}
In order to ensure that the student model learns only from meaningful and reliable predictions, we further incorporate the idea of an uncertainty aware scheme similar to \citep{yuwali2019}. This means, to estimate uncertainty, we use a Bayesian approach with Monte Carlo Dropout, applying dropout during inference to generate multiple stochastic forward passes. The variation across these passes provides an uncertainty measure where consistent predictions indicate high confidence, while variability signals uncertainty. More precisely, to estimate uncertainty, we perform $T$ stochastic forward passes on the teacher model, applying random dropout and adding Gaussian noise to each input image. The obtained probability vectors $\{\mathrm{p}_t\}_{t=1}^T$, where $\mathrm{p}_t = \modelT(T_{\phi_t}^\mathrm{P}(\Xlu))$ for some $\phi_t$, are then used to quantify uncertainty by calculating the predictive entropy, 
\begin{align*}
    \uncert = -\sum_c\left(\frac{1}{T}\sum_t \mathrm{p}_t^c\right) \log\left(\frac{1}{T}\sum_t \mathrm{p}_t^c\right) ,
\end{align*}
where $\mathrm{p}_t^c$ denotes the probability of class $c$ in the forward pass $t$.
\\\\
Let $\Tgp = \Tgpdef$ denote the perturbations applied within our transformation consistent regularization scheme, and $\mathrm{p}_\paramS$ and $\mathrm{p}_\paramT$ the probabilities obtained by $\modelS(\Tgp(\Xlu))$ and $\Tg(\modelT(\Xlu))$, respectively. Further, let $\Tilde{\uncert} = \Tg(\uncert)$ denote the estimated uncertainty map with applied geometric transformations.
The consistency loss including uncertainty is then defined as a weighted mean squared error (MSE),
\begin{align*}
    \Lcon_t(\modelS(\Tgp(\Xlu)), \Tg(\modelT(\Xlu))) = \frac{\sum_{ij}\chi(\Tilde{\uncert}_{ij}<\tau(t)) v_{ij}^\gamma\left( \mathrm{p}_\paramS - \mathrm{p}_\paramT\right)_{ij}^2}{\sum_{ij}\chi(\Tilde{\uncert}_{ij}<\tau(t)) v_{ij}^\gamma} ,
\end{align*}
where $\tau$ denotes a threshold that ensures that only the most certain pixel predictions are selected. Using the weighting function $\lambda(t)$ in (\ref{equ:rampup}), we define $\tau(t) = \ln(2)\left(\frac{3}{4} + \frac{\lambda(t)}{4} \right)$, allowing for a gradual increase in the uncertainty threshold and thus enabling the model to initially focus on the most confident predictions and progressively incorporate less confident ones as training progresses. This approach helps stabilize training early on when the model predictions might be noisy and encourages better utilization of unlabeled data as the model becomes more reliable.

\subsection{Prior knowledge}

\begin{figure}[t]
    \centering
    \begin{subfigure}{0.45\linewidth}
    \includegraphics[width=\linewidth]{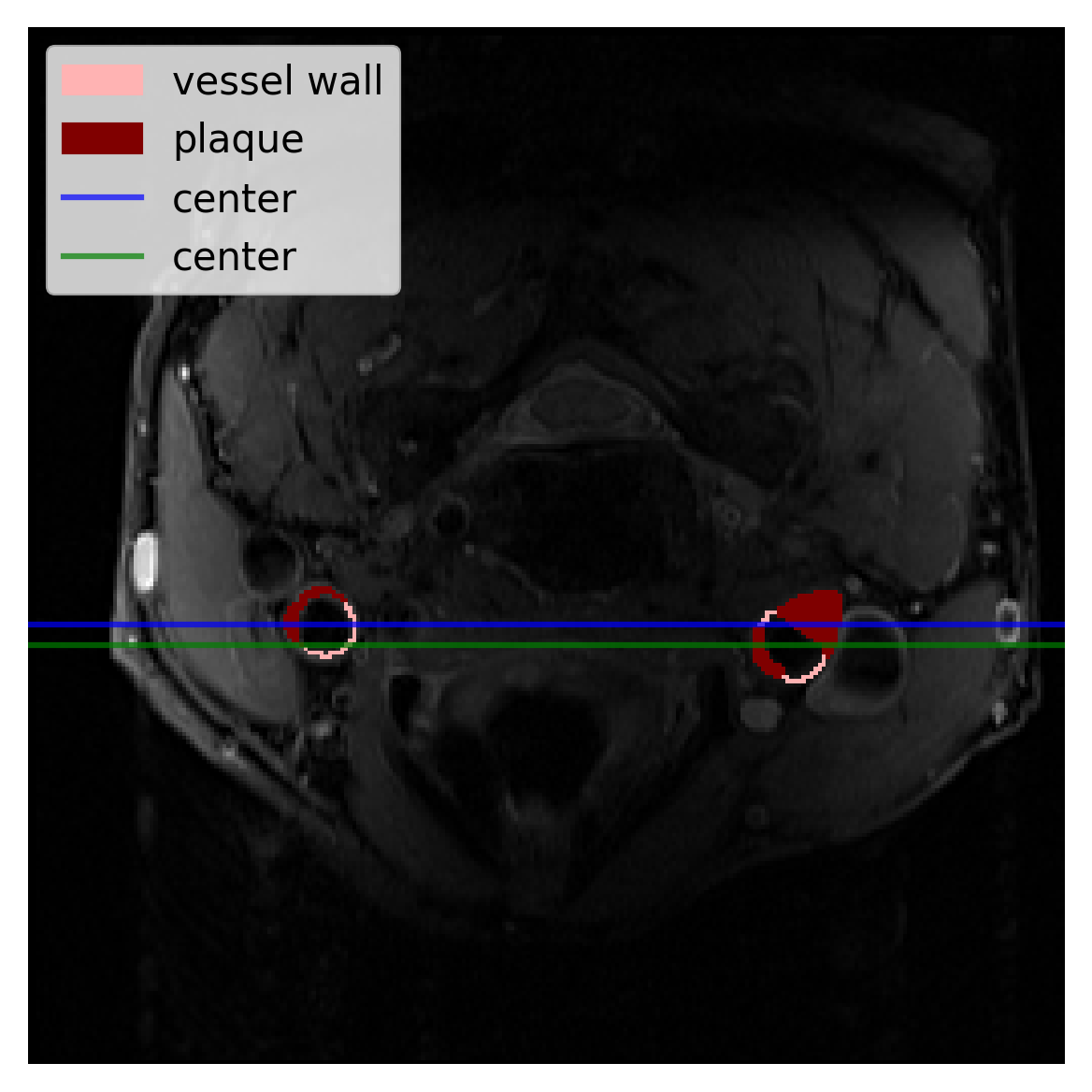}
    \caption{Patient 1} \label{p1}
    \end{subfigure}
    \hfill
    \begin{subfigure}{0.45\linewidth}
    \includegraphics[width=\linewidth]{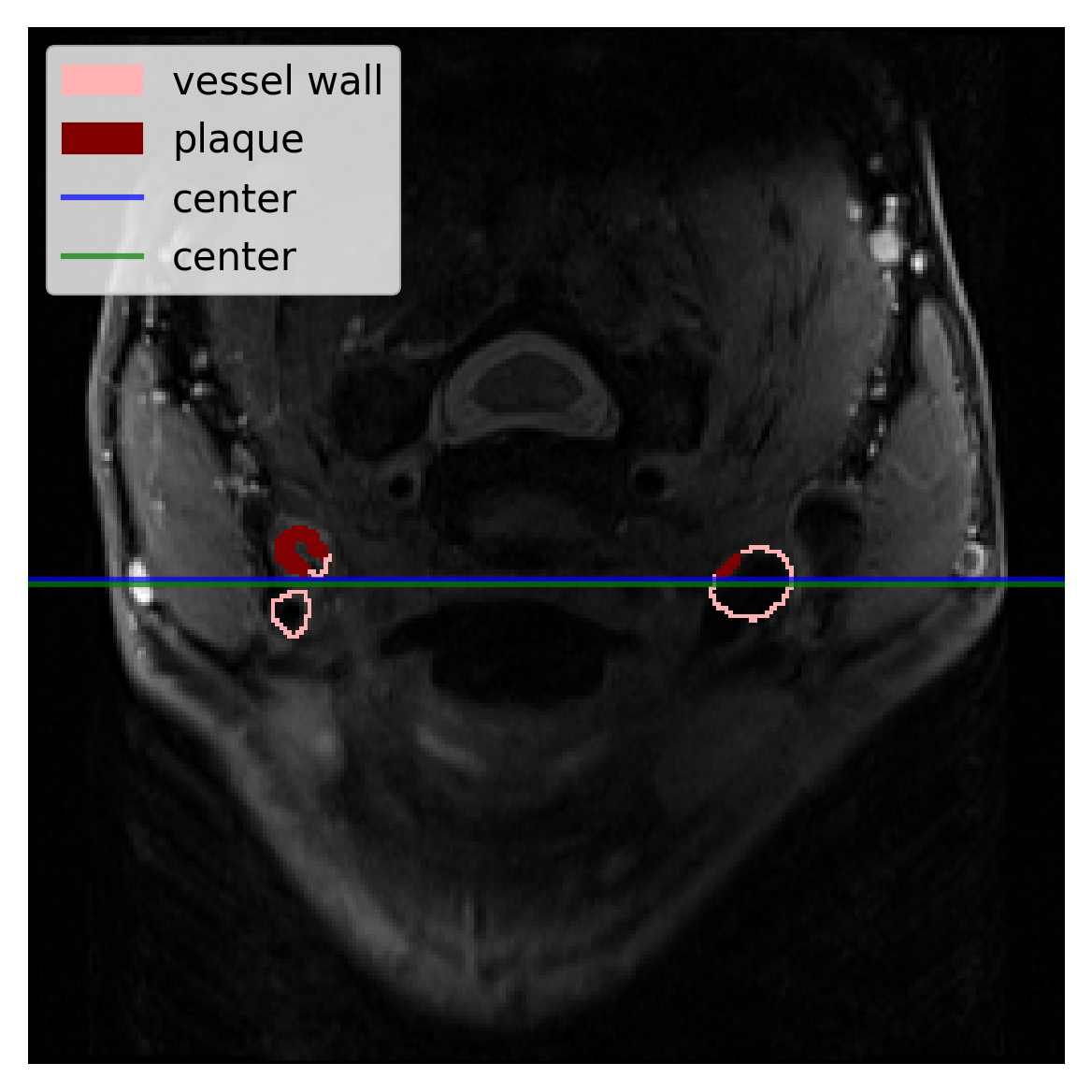}
    \caption{Patient 2} \label{p2}
    \end{subfigure}
    \caption{Left and right carotid arteries of two patients with highlighted ground-truth vessel walls and plaques. The center of each artery is highlighted by a horizontal line. 
    Both images \ref{p1} and \ref{p2} show that the carotid arteries can be highly non-symmetric, however their centers are approximately on the same horizontal line.}
    \label{fig:symmetry}
\end{figure}

For the coarse localization model, we introduce a prior 
to guide the network's predictions towards anatomically relevant regions. This prior integrates knowledge about the expected locations and number of carotid arteries, effectively constraining the model's predictions and enhancing localization accuracy. The loss function for the coarse localization network can thus be formulated as,
\begin{align}\label{equ:prior}
    \Ltotal_t(\modelS(\Xlu), \Y) = \Lsup_t(\modelS(\Xl), \Y) +\lambda(t)\Lcon_t(\modelS(\Tgp(\Xlu)), \Tg(\modelT(\Xlu))) + \omega\,\Lprior_t(\modelS(\Xl)) ,
\end{align}
where $\Lprior$ denotes the prior loss function that is applied to the predictions calculated for the supervised loss and $\omega$ a weighting coefficient. The prior function that is used within the prior loss function is also applied to the network output predictions 
and defined to ensure that the number of connected components (carotid artery vessels) found on each side of the image is restricted to a minimal number of one and a maximal number of two, reducing the false positive identification of other tissue structures. Additionally, the prior function enforces approximate symmetry by ensuring that the centers of the detected carotid arteries are approximately on the same horizontal line.
In Figure~\ref{fig:symmetry}, the left and right carotid artery vessel walls (and plaques) are displayed. The reason for enforcing only approximate symmetry of the two carotid arteries is that we observed that pathological changes, such as carotid artery plaques, and thus also the shape of carotid artery vessel walls can be highly non-symmetric, Figure~\ref{p1}. Further, also the bifurcation of the carotid artery may not always be symmetrical, Figure~\ref{p2}. For this, we constructed a geometric prior that accounts for natural asymmetries by focusing on positional alignment rather than shape constraints.

%%%%%%%%%%%%%%
\section{Experiments and Results}\label{sec:experiments}
%%%%%%%%%%%%%%

\subsection{Experimental Settings}

In the following, we introduce the dataset used within our experiments and provide all the details on the experimental set-up.

\paragraph*{Dataset}
The clinical dataset used for the evaluation of the proposed segmentation model consisted of 52 patients. This study was approved by the local ethics committee and written informed consent from all patients was obtained. For each patient, five MRI sequences (T1w, T1ce, T2w, TOF, PDw) were acquired with the following parameters:
\begin{itemize}
    \item PDw Turbo-Spin-Echo sequence, fat saturated: TR = $1290$~ms, TE = $30$~ms, turbo factor = $7$, rephasing flip angle = $172^{\circ}$, FOV = $140 \times 140$~mm, acquisition matrix = $256 \times 205$, pixel spacing: $0.55 \times 0.55$~mm, slice orientation: transversal, slice thickness: $2$~mm, spacing between slices: $2.2$~mm
    \item T1w Turbo-Spin-Echo sequence, fat saturated and T1ce Turbo-Spin-Echo sequence, fat saturated: TR = $639$~ms, TE = $12$~ms, turbo factor = $2$, rephasing flip angle = $144^{\circ}$, FOV = $140 \times 140$~mm, acquisition matrix = $256 \times 179$, pixel spacing: $0.55 \times 0.55$~mm, slice orientation: transversal, slice thickness: $2$~mm, spacing between slices: $2.2$~mm
    \item T2w Turbo-Spin-Echo sequence, fat saturated, fat saturated: TR = $1290$~ms, TE = $61$~ms, turbo factor = $7$, rephasing flip angle = $172^{\circ}$, FOV = $140 \times 140$~mm, acquisition matrix = $256 \times 205$, pixel spacing: $0.55 \times 0.55$~mm, slice orientation: transversal, slice thickness: $2$~mm, spacing between slices: $2.2$~mm
    \item TOF 3D sequence with multiple overlapping slabs: TR = $21$~ms, TE = $3.6$~ms, flip angle = $25^{\circ}$, TONE pulse = $50\%$, slice orientation = transversal, slice thickness = $1$~mm, FOV = $149 \times 199$~mm, acquisition matrix = $384 \times 187$, pixel spacing = $0.52 \times 0.52$~mm
\end{itemize}

The sequences were acquired by a Siemens Verio 3 T MRI system (Siemens Healthineers, Erlangen, Germany) in the years 2011 to 2015. Each sequence consisted of multiple 2D slices, representing 2D cross-sections of the neck. For 29 of the 52 patients, the PDw sequence was partially annotated by two independent experts of the local radiological department using ITK-Snap version 4.2\footnote{\url{http://www.itksnap.org/pmwiki/pmwiki.php?n=Main.HomePage}, Accessed 21.01.2015.}. This sequence was selected for labeling due to its high contrast, which allows for better differentiation between the carotid artery vessel wall, lumen and plaques. 
To train the localization network, we required symmetrically labeled image slices, meaning both the left and right carotid arteries had to be annotated. As a result, 19 out of the 29 labeled patients were suitable for coarse localization, providing a total of 216 image slices. For fine segmentation, we used 529 extracted ROI slices representing 39 carotid arteries across 24 patients. Additionally, 458 unlabeled slices (corresponding to 916 ROI slices from 23 patients) were incorporated into the semi-supervised training. Five symmetrically labeled patients were included in coarse localization but excluded from fine segmentation due to insufficient annotation quality.
\\

\paragraph*{Evaluation Metrics}
To evaluate the model performance several metrics, including Dice Similarity Coefficient (DSC), Intersection over Union (IoU), precision, and recall, were calculated, comparing the predicted segmentation masks with expert-labeled ground truths. 
These metrics were computed for each class - background, vessel wall, and plaque - and subsequently averaged. To complement these quantitative measures, we additionally conducted an expert based assessment of model performance to validate the accuracy of the obtained segmentations.
\\

\paragraph*{Preprocessing}
Multi-modal image analysis has shown to be superior to single-modal image analysis since the combination of different modalities can achieve complementary information between images, \citep{li2023carotid, zhu2022complex}. However, performing multi-modal analysis still remains a challenge, particularly when image labels are available only for a subset of the sequences. In our case, only one of the five different MRI sequences was partially annotated by experts using ITK-Snap. This means, segmentation masks were provided only for one sequence and a subset of the slices. Additionally, the MRI sequences were unaligned, meaning that the labeled masks were valid only for the sequence that was actually annotated. To address this issue, all MRI sequences were first resampled to the labeled PDw sequence to ensure matching resolutions. For spatial alignment, image registration was performed using rigid and scaling transformations. Further, $N4$-bias field correction was applied to the MRI sequences to reduce intensity inhomogeneities caused by variations in the magnetic field of the MRI scanner.
Manual preprocessing steps, such as removing black slices or those with structural distortions caused by resampling or registration were also performed to improve the overall quality of the data.
\\

\paragraph*{Implementation Details}
The experiments were conducted using PyTorch v2.3.0 and executed on a Nvidia RTX A5000 GPU. For the preprocessing of the MRI sequences we used the Python libraries ANTsPy and NiBabel. Following registration, the slices were standardized by subtracting the mean and dividing by the standard deviation of each sequence, and normalized by scaling their pixel-values to the range $\left[ 0,1\right]$. For our experiments, the labeled slices were divided patient-wise into training, validation, and test sets. For the localization task, $80\%$ of the patients were used for training, $10\%$ for validation, and $10\%$ for testing. Fine segmentation employed a 5-fold cross-validation strategy across patients with ground-truth annotations, where a fixed but randomly selected subset of patients was reserved for validation in each fold. Unlabeled slices were incorporated into the semi-supervised training for both localization and segmentation tasks. The reported results represent the mean evaluation metrics across all folds. 
For training, we used ADAM optimizer with an initial learning rate of $10^{-4}$ for localization and $10^{-5}$ for fine segmentation, in combination with a scheduler that reduces the learning rate by a factor of $0.1$ if no improvement is made within $10$ epochs. Further, a weight decay ($L_2$ regularization) of $0.0001$ for localization and $0.0005$ for fine segmentation was used within ADAM. The momentum parameters of ADAM, $\beta_1$ and $\beta_2$, were set to $0.9$ and $0.999$, respectively. We decided to train the models for $120$ epochs and terminated the training if the segmentation results on the validation set did not improve over 20 and 40 epochs for coarse localization and fine segmentation, respectively. 
During training, data augmentation was dynamically applied to labeled data in order to expand the training set, including random flipping and rotations, using the Albumentations library \citep{buigkh2020}. For the labeled data, we further included image slices that were extracted using slightly shifted bounding boxes to obtain additional data with non-centered carotid arteries. Regarding the model architecture, the number of encoder and decoder blocks within the fine segmentation model was set to 4 and within the coarse localization model to 5. For both models we used a kernel size of $3\times 3$ within convolutional layers and a kernel size of $2\times 2$ for downsampling convolutions. Further, a dropout rate of $0.3$ was applied in the encoder layers of both models. The size of the input images for the coarse localization was $\dimgh\times\dimgw = 256\times 256$. For the fine segmentation we used ROI images of size $\dimgh\times\dimgw = 64\times 64$.

For semi-supervised learning, the contribution of the consistency loss was gradually increased over the first $R=60$ (for localization) and $R=40$ (for fine segmentation) epochs using the weighting function $\lambda(t)$ defined in equation~(\ref{equ:rampup}), ensuring that the model initially focuses on learning from labeled slices and slowly increases the contribution of learning from unlabeled slices. The factor $k$ within the weighting function $\lambda(t)$ was set to $k=10$ and $k=20$ for coarse localization and fine segmentation, respectively.
The smoothing coefficient used within the EMA update of the teacher weights was set to $\alpha = 0.999$.
For the weighting coefficient $\lambda_\mathrm{loc}$ within the supervised loss of the coarse localization we used $\lambda_\mathrm{loc} = 0.5$ and for $\delta$, controlling the relative weighting of positive and negative examples, we used $\delta = 0.7$. Within the supervised loss of the fine segmentation we selected $\lambda_\mathrm{seg} = 0.5$ and $\delta = 0.6$ as suggested by \citep{yesasc2022}, and set $\vartheta = 0.25 $ .
For uncertainty estimation, we performed $T=8$ stochastic forward passes to calculate the predictive entropy. The weight for the prior loss within the coarse localization loss in (\ref{equ:prior}) was set to $\omega = 0.1$ and approximate symmetry was defined such that the maximal difference between the $y$-coordinates of the centers of the left and right carotid arteries has to be $\leq20$.

To ensure a fair comparison between different models and learning strategies, we maintained consistent experimental settings, including identical loss functions, hyperparameters, and preprocessing steps across all experiments.

\subsection{Experimental Results}

Experiments were performed for coarse localization and fine segmentation, respectively. As semi-supervised approach for the coarse localization and different design choices within the experiments for fine segmentation, we followed the one-way consistency approach introduced in Section~\ref{sec:methods}. In order to analyze the contribution of semi-supervised learning in general and semi-supervised learning with and without uncertainty estimation, additional experiments were conducted.

The first experiments were performed for coarse localization in order to identify and extract the ROI images used within the subsequent fine segmentation step. Following the implementation details above, the coarse localization network has shown to correctly identify left and right carotid artery in 788 out of 789 image slices ($99.87 \%$). This performance was evaluated by two expert radiologists. The center points of the correctly identified carotid arteries were extracted and used to crop each sequence to ROI image slices of size $64\times 64$. Within the image slice which was not correct, both carotid arteries were detected however also some other tissue structure was classified as foreground, leading to a shift of one carotid center point and thus to some plaque pixel outside the bounding box. Using no prior, i.e.~no knowledge on the number of detected connected components and approximate symmetry, led to missed or cut-off carotid arteries in 38 slices, leading to a decreased detection rate of $95.18\%$ (see Table~\ref{tab:prior_vs_no}).

\begin{table}[t]
    \centering
    \begin{tabular}{l|c c}
         &  failed detection & detection rate\\
         \hline
        no Prior & 38/789 slices & $95.18\%$ \\
        Prior & 1/789 slices & $99.87 \%$
    \end{tabular}
    \caption{Number of slices for which one or both carotid arteries have not been correctly identified with and without incorporating prior knowledge.} 
    \label{tab:prior_vs_no}
\end{table}

\begin{table*}[t]
    \centering
    \begin{tabular}{l|p{1.7cm}  p{1.7cm} p{1.7cm} p{1.7cm}}
     & Dice & IoU & Precision & Recall \\
     \hline
     Basic U-Net input fusion &  0.8265 &  0.7350 &  0.7772 & 0.9074 \\
     Basic U-Net bottleneck fusion &  0.8702 & 0.7897 & 0.8339 & \textbf{0.9200}\\
     Our U-Net input fusion & 0.8549 & 0.7706 & 0.8250 &  0.8950 \\
     Our U-Net bottleneck fusion & \textbf{0.8725} & \textbf{0.7930} & \textbf{0.8475} & 0.9040 \\
    \end{tabular}
    \caption{Comparison of basic U-Net and our U-Net using different fusion points for binary segmentation.}
    \label{tab:late_early_binary}
\end{table*}

\begin{figure}[t]
    \centering
    \includegraphics[width=0.8\linewidth]{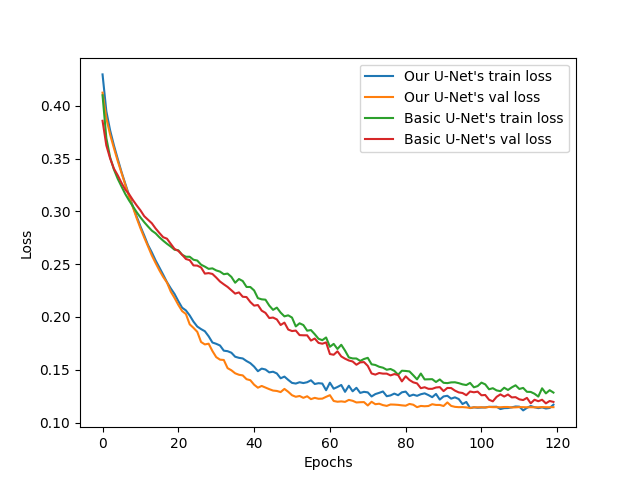}
    \caption{Performance of our U-Net and basic U-Net: Train and validation loss versus number of epochs.}
    \label{fig:performance}
\end{figure}

Given the ROI images containing the carotid arteries, several design choices of the fine segmentation module were evaluated. First, we aimed to demonstrate the effectiveness of the fusion strategy in both the standard U-Net architecture and our multi-sequence version of U-Net introduced in Section~\ref{sec:methods}. The basic U-Net architecture in combination with input fusion not only represents a baseline model but also refers to the model architecture used in \citep{wayu2024}, which has shown superior performance to all existing published methods for carotid plaque segmentation in multi-sequence MRI, and builds upon a U-Net with doubled filters in the encoder.
Within both model architectures different fusion points (fusion at input level and fusion in the bottleneck layer) are reported. The predictions were made and evaluated slice-wise. Table~\ref{tab:late_early_binary} shows a quantitative comparison of the two models trained and evaluated for binary segmentation of the carotid artery (vessel wall and plaque versus background) given five different MRI sequences. For basic U-Net, bottleneck fusion provided significant improvements of over $4$ percentage points in Dice, IoU, and precision compared to input fusion. Our proposed U-Net architecture led to more similar results for input and bottleneck fusion with a total difference of about $2$ percentage points with respect to Dice, IoU and precision in favor of bottleneck fusion. Compared to input fusion, our modifications to the basic U-Net architecture led to substantial performance gains, demonstrating that our proposed model is able to effectively capture and utilize information from multi-channel input data.
In summary, our U-Net model achieved the overall best Dice, IoU and precision scores and provided more robust segmentation results as the difference between input and bottleneck fusion became less severe than in the U-Net case. U-Net with bottleneck fusion provided the best recall scores, indicating that our U-Net model provided fewer false positives compared to basic U-Net, while basic U-Net found more true positives. In other words, predictions from our model contain fewer true positives than the ones from the basic U-Net architecture while there are way less false positives found by our model. Note that, the behavior of our model and basic U-Net was consistent with the findings in \citep{ayhu2018}, which found that late fusion achieved better results. In addition to improved segmentation performance, our proposed U-Net also outperformed the basic U-Net in terms of training efficiency (see Figure~\ref{fig:performance}). Specifically, it exhibited faster convergence and consistently achieved lower training and validation losses across all epochs and folds.

\begin{figure}[t]
    \centering
    \begin{subfigure}{.9\linewidth}
    \includegraphics[width=\linewidth, trim = 0 50 0 20, clip]{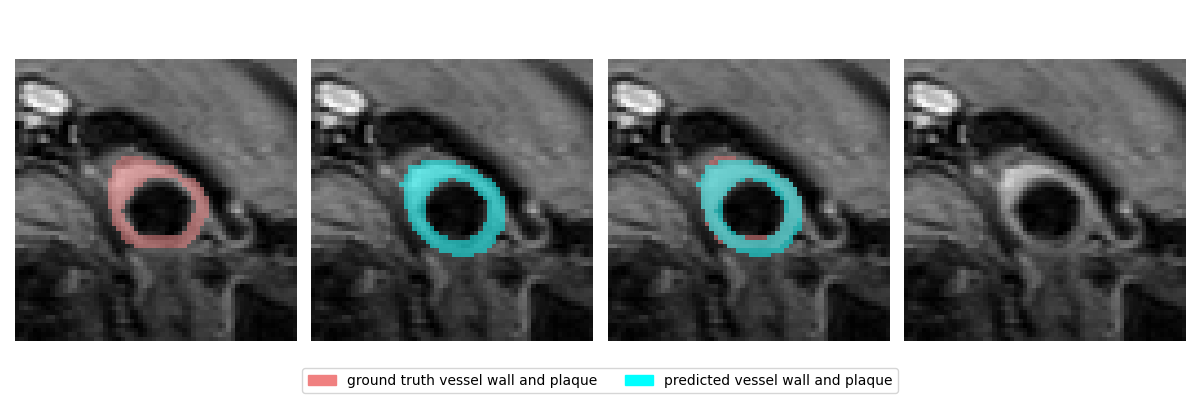}
    \end{subfigure}
    \begin{subfigure}{.9\linewidth}
    \includegraphics[width=\linewidth, trim = 0 50 0 20, clip]{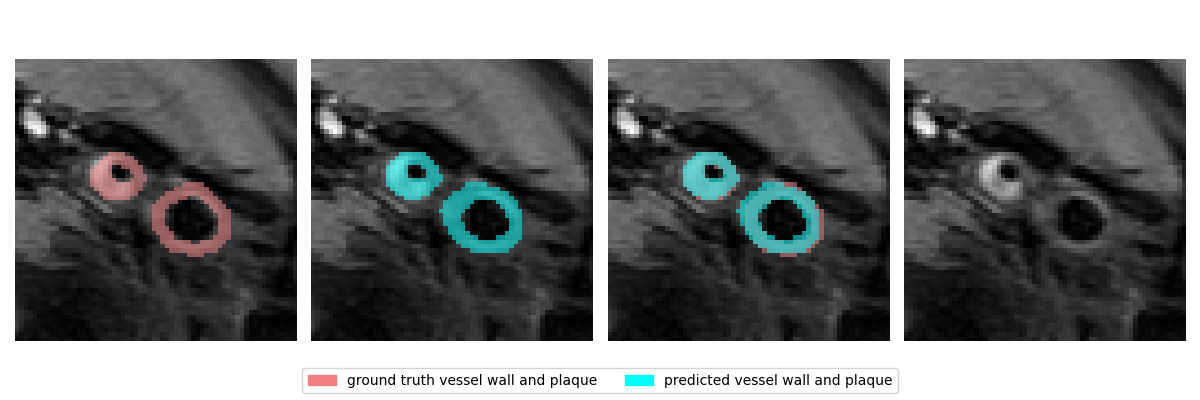}
    \end{subfigure}
    \begin{subfigure}{.9\linewidth}
    \includegraphics[width=\linewidth, trim = 0 50 0 20, clip]{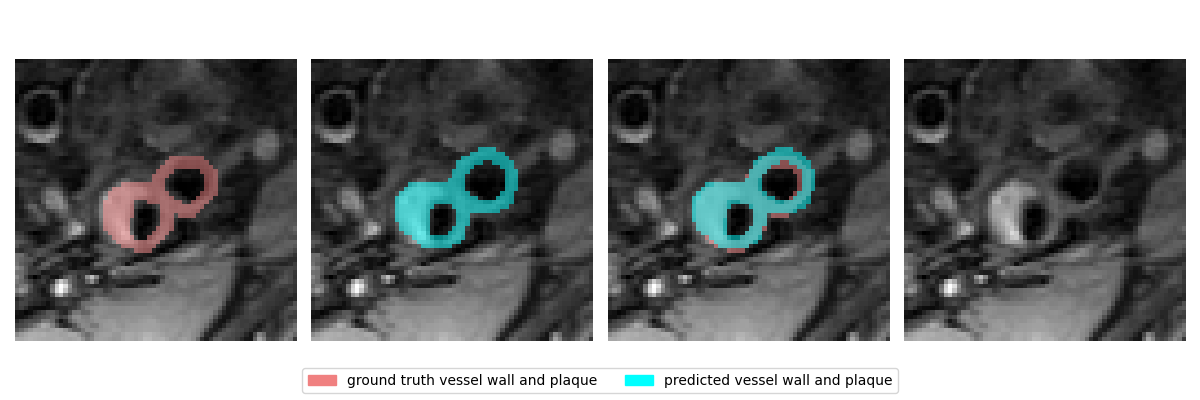}
    \end{subfigure}
    \begin{subfigure}{.9\linewidth}
    \includegraphics[width=\linewidth, trim = 0 50 0 20, clip]{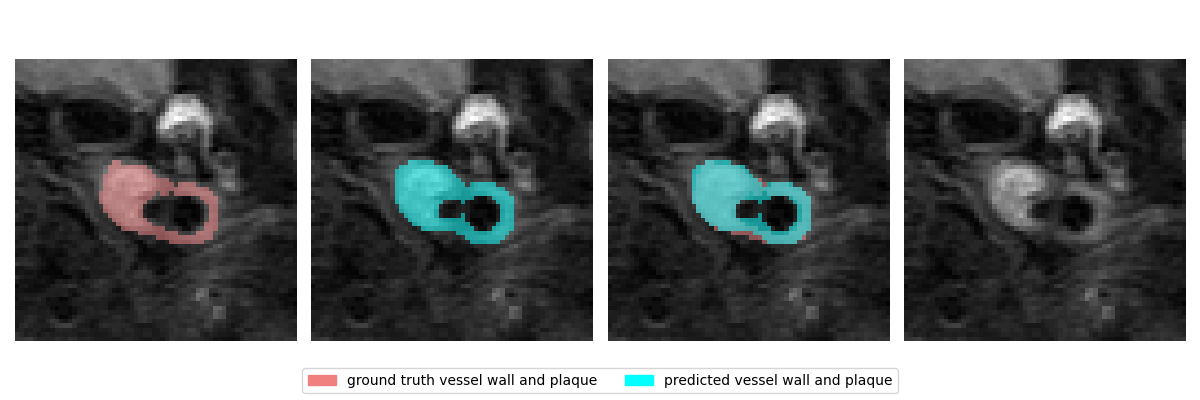}
    \end{subfigure}
    \begin{subfigure}{.9\linewidth}
    \includegraphics[width=\linewidth, trim = 100 0 100 255, clip]{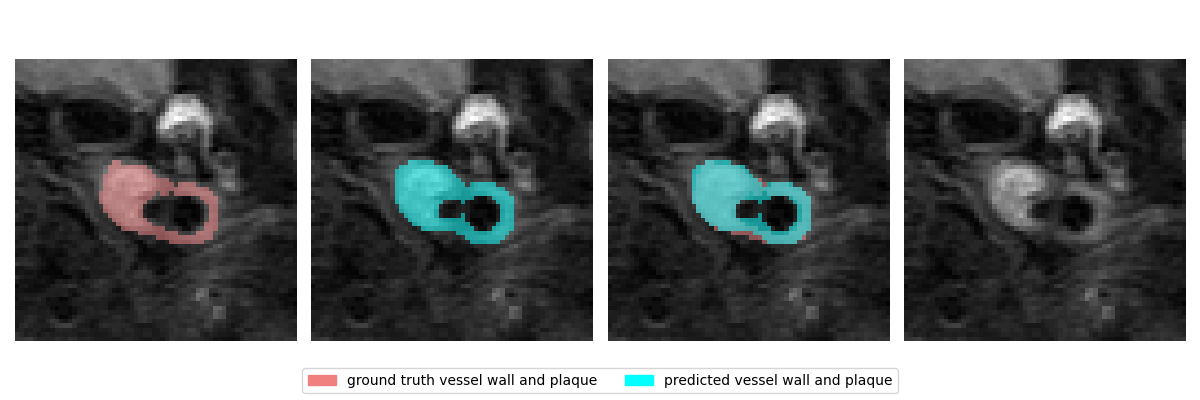}
    \end{subfigure}
    \caption{Results of binary carotid artery vessel wall and plaque segmentation using our U-Net and late fusion. From left to right: Ground truth mask, predicted mask, their overlay, PDw image slice.}
    \label{fig:example_output}
\end{figure}

In Figure~\ref{fig:example_output} we present exemplary outputs of our U-Net model in combination with bottleneck fusion for the binary case in comparison with the ground truth mask. The model was trained using image slices from 16 patients, while slices from 4 additional patients were used for validation and another 4 for testing. The predicted segmentation masks demonstrate a strong correspondence with the ground truth annotations provided by radiologists.
In particular, the predicted masks accurately capture the target regions with minimal deviations, effectively delineating the carotid artery vessel wall and plaque. Not only do they successfully identify the intended structures, but they also align more closely with the underlying images. For instance, in the last example of Figure~\ref{fig:example_output}, the model correctly separates the two arteries, highlighting its accuracy.

\begin{figure}[t]
    \centering
    \begin{subfigure}{.9\linewidth}
    \includegraphics[width=\linewidth, trim = 0 50 0 30, clip]{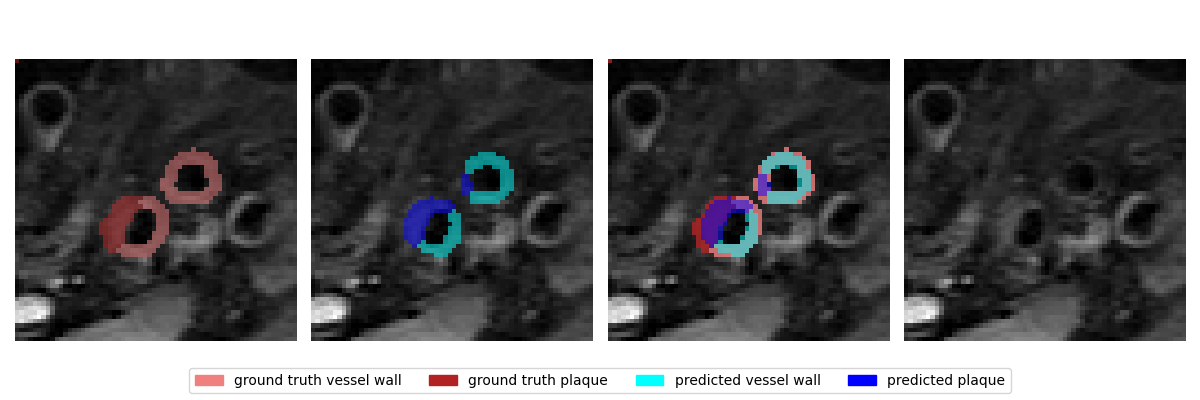}
    \end{subfigure}
    \begin{subfigure}{.9\linewidth}
    \includegraphics[width=\linewidth, trim = 0 50 0 30, clip]{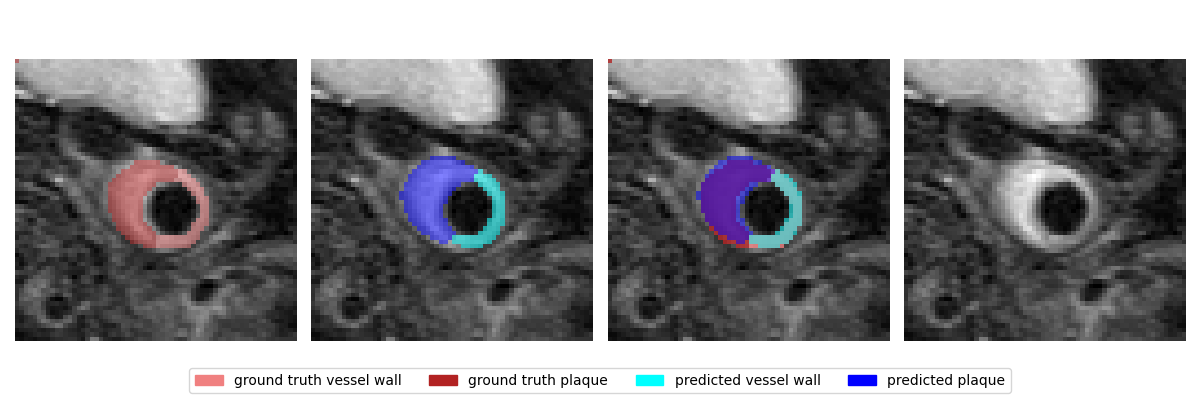}
    \end{subfigure}
    \begin{subfigure}{.9\linewidth}
    \includegraphics[width=\linewidth, trim = 0 50 0 30, clip]{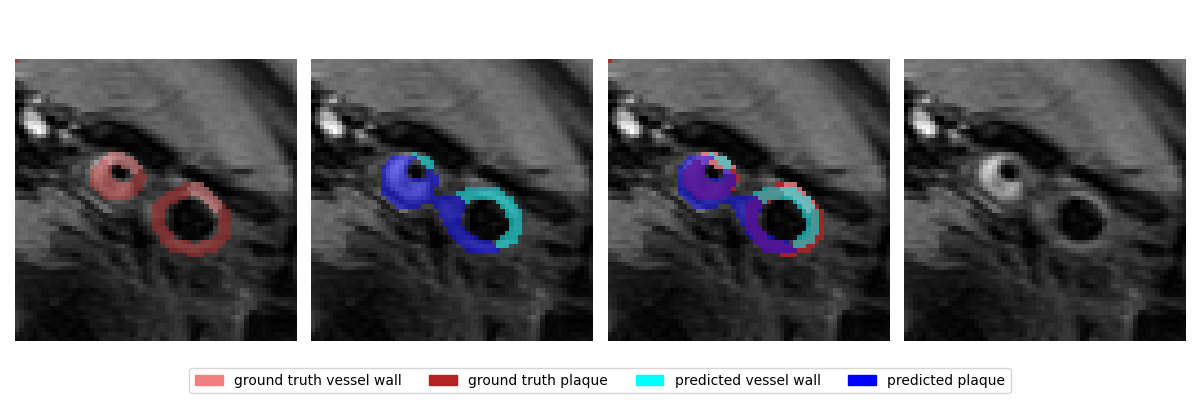}
    \end{subfigure}
    \begin{subfigure}{.9\linewidth}
    \includegraphics[width=\linewidth, trim = 0 50 0 30, clip]{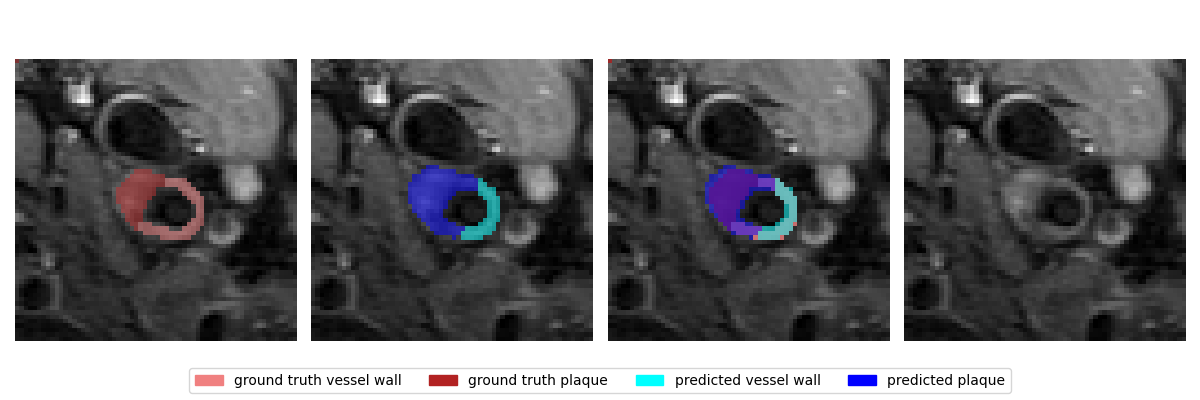}
    \end{subfigure}
    \begin{subfigure}{.9\linewidth}
    \includegraphics[width=\linewidth, trim = 100 0 110 255, clip]{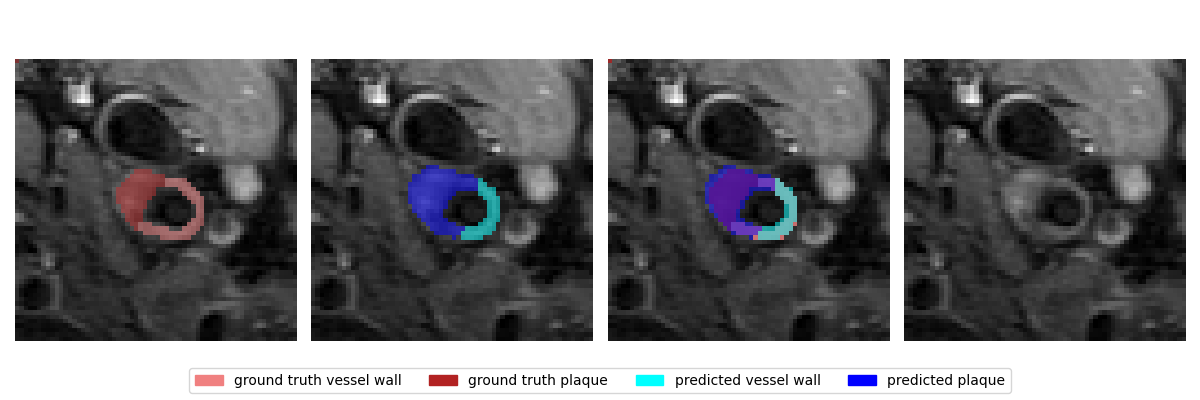}
    \end{subfigure}
    \caption{Results of multiclass carotid artery vessel wall and plaque segmentation using our U-Net and late fusion. From left to right: Ground truth mask, predicted mask, their overlay, PDw image slice.}
    \label{fig:example_output_multiclass}
\end{figure}

\begin{table*}[t]
    \centering
    \begin{tabular}{l|p{1.7cm}  p{1.7cm} p{1.7cm} p{1.7cm}}
     & Dice & IoU & Precision & Recall \\
     \hline
     Basic U-Net input fusion &  0.6761 & 0.5604 & 0.6535 & 0.7226 \\
     Basic U-Net bottleneck fusion &  \textbf{0.7010} & \textbf{0.5864} & \textbf{0.7441} & \textbf{0.7290} \\
     Our U-Net input fusion & 0.6753 &  0.5617 &  0.6725 & 0.7157 \\
     Our U-Net bottleneck fusion & 0.6787 &  0.5672 &  0.7146 &  0.7080\\ 
    \end{tabular}
    \caption{Comparison of basic U-Net and our U-Net for multiclass segmentation.}
    \label{tab:late_early_multiclass}
\end{table*}

For multiclass segmentation, Figure~\ref{fig:example_output_multiclass} displays representative output masks generated by our U-Net model using late fusion, where again image slices from 16 patients were used for training, slices from 4 patients for validation and slices form 4 patients for testing. 
Similar to the binary case, we observed a good coverage of the predicted masks and the ground-truth annotations provided by radiologists. Again, our U-Net model often seems to reflect the structure of the vessel more accurately compared to the ground truth mask.
 
\begin{figure}[t]
    \centering
    \begin{subfigure}{.9\linewidth}
    \includegraphics[width=\linewidth, trim = 0 250 0 0, clip]{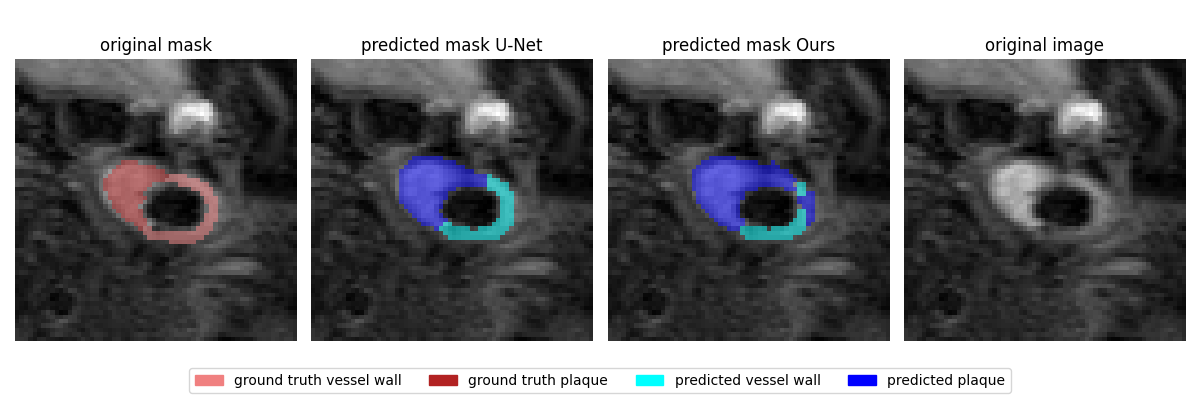}
    \end{subfigure}
    \begin{subfigure}{.9\linewidth}
    \includegraphics[width=\linewidth, trim = 0 40 0 40, clip]{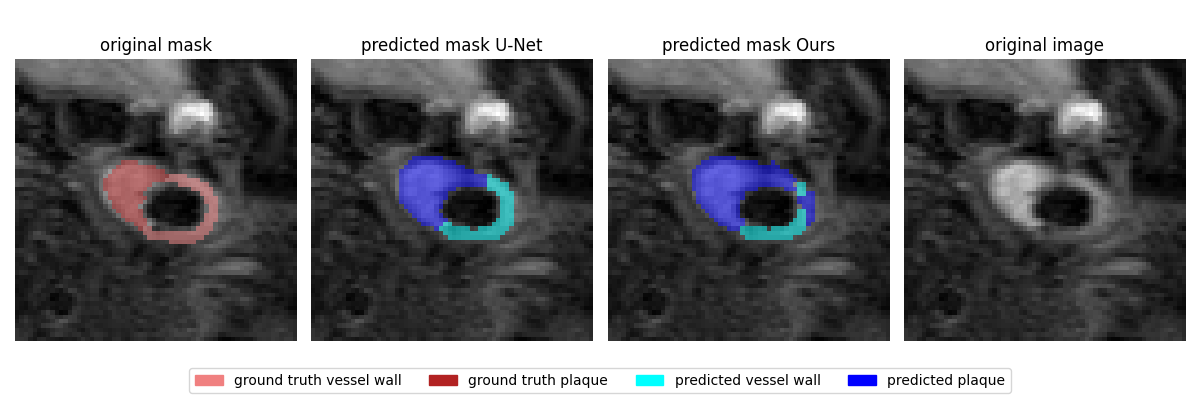}
    \end{subfigure}
    \begin{subfigure}{.9\linewidth}
    \includegraphics[width=\linewidth, trim = 0 40 0 40, clip]{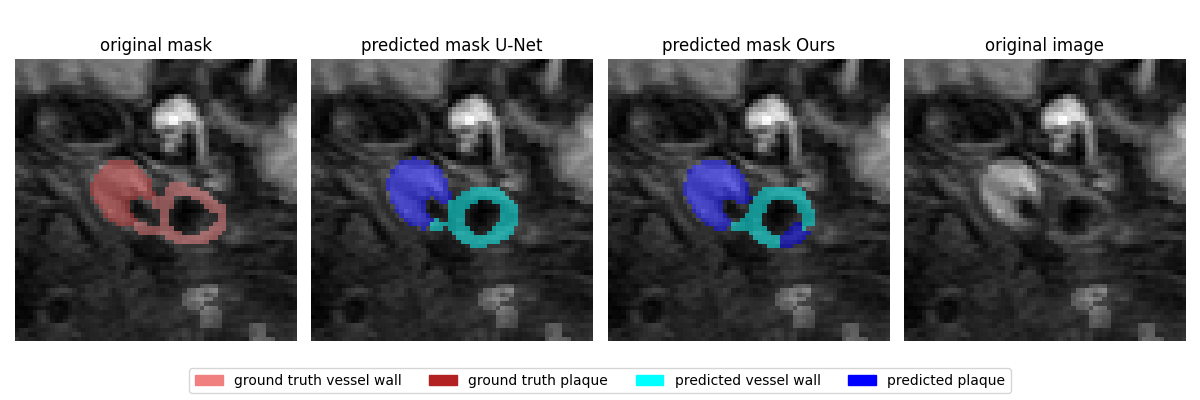}
    \end{subfigure}
    \begin{subfigure}{.9\linewidth}
    \includegraphics[width=\linewidth, trim = 0 40 0 40, clip]{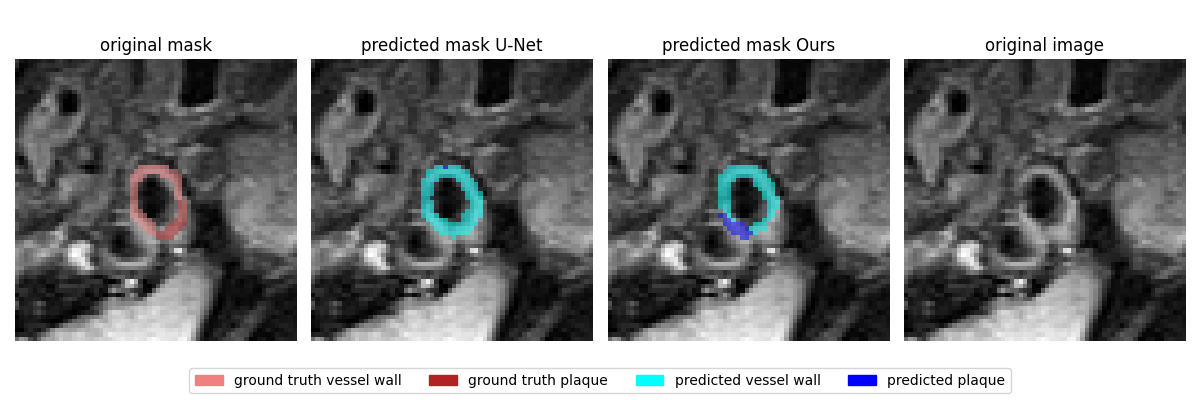}
    \end{subfigure}
    \begin{subfigure}{.9\linewidth}
    \includegraphics[width=\linewidth, trim = 0 40 0 40, clip]{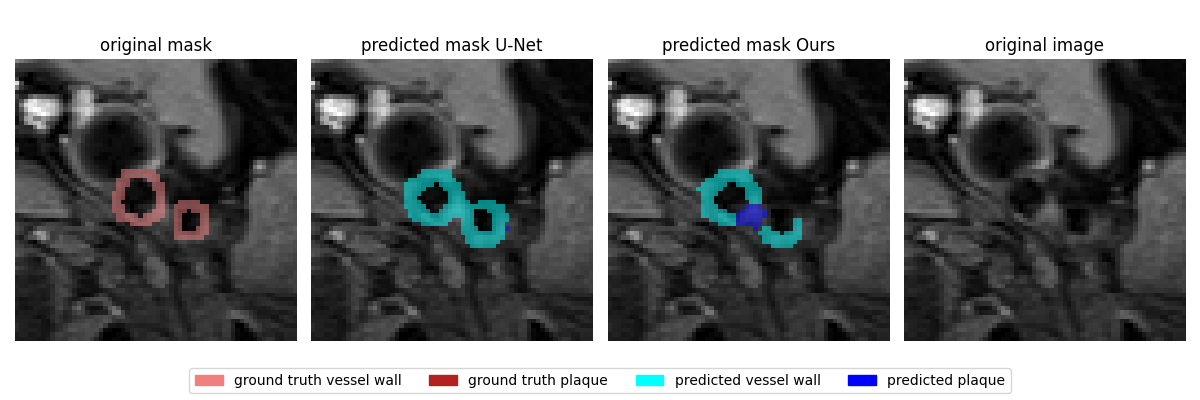}
    \end{subfigure}
    \begin{subfigure}{.9\linewidth}
    \includegraphics[width=\linewidth, trim = 100 0 100 255, clip]{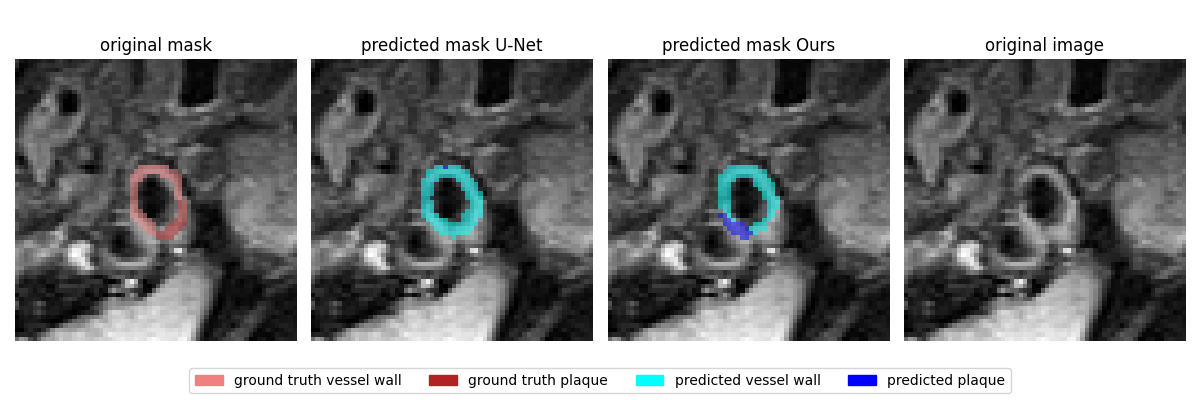}
    \end{subfigure}
    \caption{Comparison of multiclass ground truth masks, predicted masks using basic U-Net with late fusion, and predicted masks using our U-Net with late fusion.}
    \label{fig:multiclass_unet_vs_marie}
\end{figure}

We also extended our analysis of the fusion strategy to the multiclass segmentation task, where each pixel was assigned to either background, vessel wall or plaque. 
Table~\ref{tab:late_early_multiclass} presents the segmentation scores of our U-Net model and a basic U-Net model using bottleneck and input fusion. For our U-Net, input and bottleneck fusion provided similar results, underscoring the robustness under the fusion of sequences. The basic U-Net model together with bottleneck fusion was superior over input fusion and over our model architecture in contrast to the binary case. 
In order to better understand this observational mismatch of binary and multiclass segmentation, we visually analyzed predictions from both basic and our U-Net using late fusion. Figure~\ref{fig:multiclass_unet_vs_marie} displays some exemplary outputs.
The segmentation masks from the basic U-Net match the ground truth labels more accurately, while our model misclassified plaque according to the ground truth masks. However, the \textit{wrong} plaque regions from our predictions show a thickened artery wall and high intensities. This suggests that the identified plaque regions not necessarily indicate a misclassification but may be the result of inaccurate ground truth labels, which would explain the worse segmentation results of our U-Net model.

When comparing the segmentation results of Table~\ref{tab:late_early_binary} and Table~\ref{tab:late_early_multiclass}, the scores of the multiclass experiments are lower compared to the binary case. This is most likely due to misclassifications between vessel wall and plaque of both the basic U-Net and our U-Net. As discussed previously, another possible cause for the low segmentation scores are erroneous ground-truth masks which did not separate vessel wall and plaque correctly. To underpin our claim on mistakes not only in the predictions but also in the labels, we trained our U-Net model in a multiclass setting and evaluated the model in a binary setting using 5-fold cross validation. We ended up with an averaged Dice score of $0.8569$, an IoU score of $0.7728$, a precision of $0.8367$, and recall of $0.8820$. The segmentation scores are slightly lower in comparison to the binary case (see Table~\ref{tab:late_early_binary}) however,  the model was trained on a harder task with more classes to separate. Nevertheless, the scores indicate that our U-Net model with bottleneck fusion effectively distinguishes vessel wall and plaque from the background. The decrease in segmentation scores is primarily due to mismatched plaque and vessel wall pixels between the predicted and ground-truth masks.

\begin{figure}[th]
    \centering
    \begin{subfigure}{.9\linewidth}
    \includegraphics[width=\linewidth, trim = 0 50 0 20, clip]{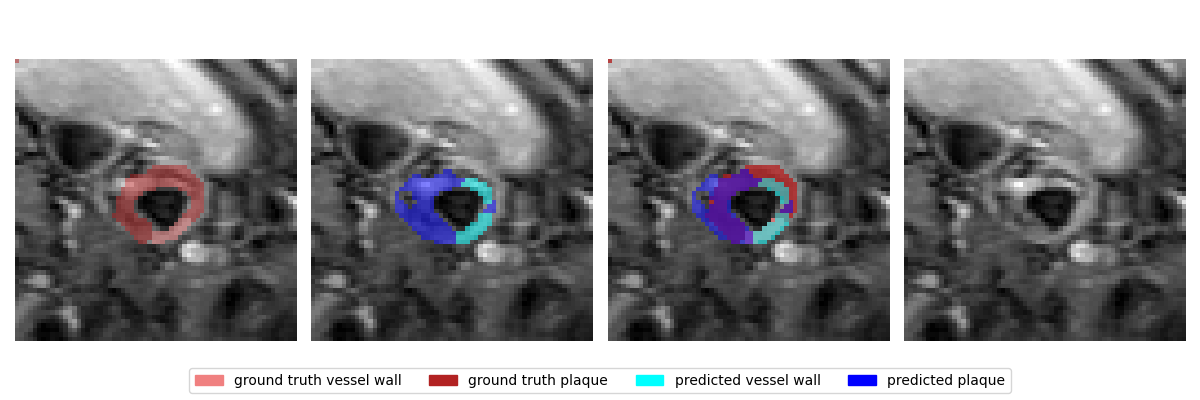}
    \end{subfigure}
    \begin{subfigure}{.9\linewidth}
    \includegraphics[width=\linewidth, trim = 0 50 0 20, clip]{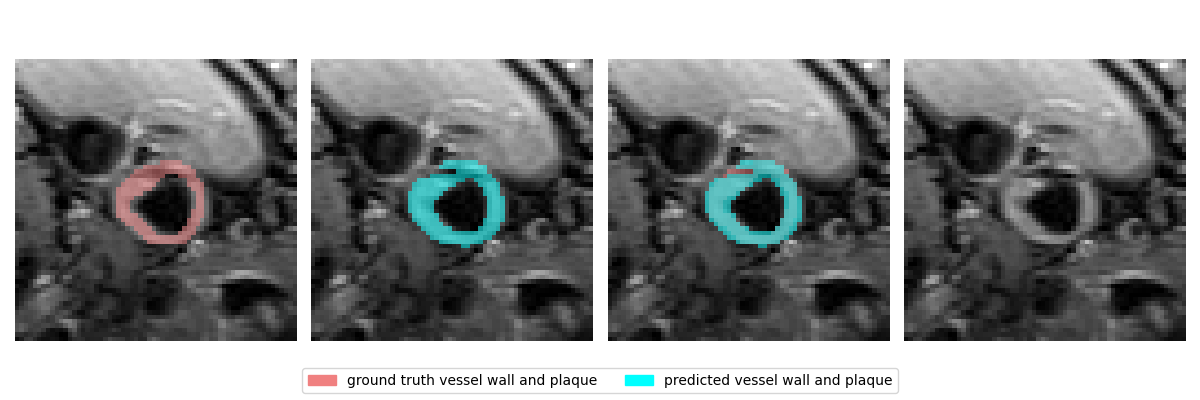}
    \end{subfigure}
    \begin{subfigure}{.9\linewidth}
    \includegraphics[width=\linewidth, trim = 100 0 100 255, clip]{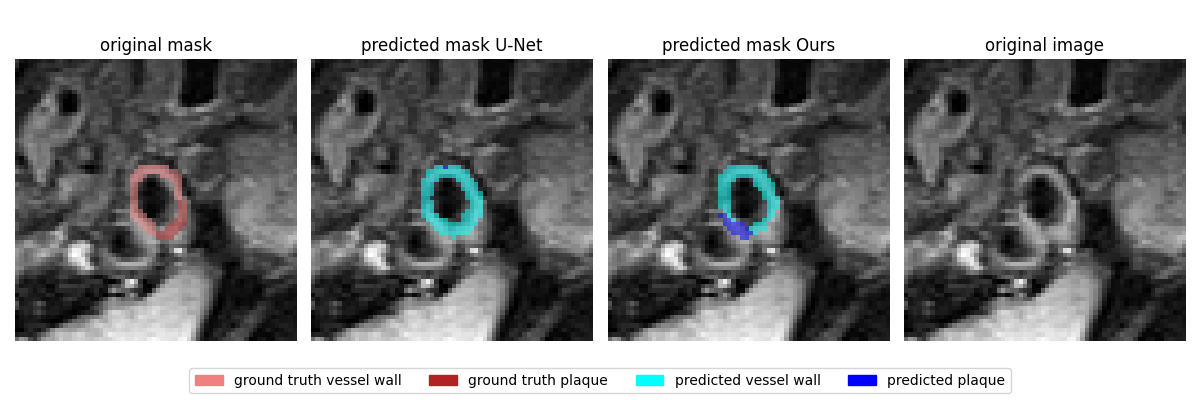}
    \end{subfigure}
    \caption{Demonstration of accuracy and integration of underlying image structures of our prediction. From left to right: Ground truth mask, predicted mask, their overlay, PDw image slice.} \label{fig:high_prec_low_recall}
\end{figure}

Figure~\ref{fig:high_prec_low_recall} further shows that our U-Net is more cautious at predicting plaque or vessel wall and reacts to the structure (change) in the underlying sequences. Representatively, this gives evidence for the low recall and high precision scores that the predicted segmentation mask of our U-Net model had.

We also aimed to analyze the impact of the semi-supervised learning scheme. Therefore, we evaluated the results of using no semi-supervision, one-way-consistency-semi-supervised learning (OWC-SSL) with a perturbed student and clean teacher model, and OWC-SSL with uncertainty estimation, together with our proposed U-Net model, late fusion, and binary segmentation. 
Table~\ref{tab:adaptions-study-semi-sup} illustrates that using semi-supervised learning with one way consistency improves Dice, IoU and recall scores compared to fully supervised training. 
The Dice, IoU and precision scores slightly improved by additionally incorporating estimations on the uncertainty of the predictions. The recall scores on the other hand slightly decreased when including estimations on the uncertainty, again indicating that there are less pixels identified as carotid artery vessel wall or plaque, but if so, it is likely to be correctly classified. A similar behavior can also be observed in the multiclass setting, reported in Table~\ref{tab:adaptions-study-semi-sup-multiclass}. 

\begin{table*}[t]
    \centering
    \begin{tabular}{l|p{1.7cm}  p{1.7cm} p{1.7cm} p{1.7cm}}
     & Dice & IoU & Precision & Recall \\
         \hline
         Our U-Net, no semi-supervision & 0.8645 & 0.7829 & \textbf{0.8608} &  0.8688\\
         Our U-Net, OWC-SSL & 0.8725 & 0.7930 & 0.8475 & \textbf{0.9040}\\
         Our U-Net, OWC-SSL + uncertainty & \textbf{0.8773} & \textbf{0.7993} &  0.8585 &  0.9002
\\ 
    \end{tabular}
    \caption{Ablation study on semi-supervised learning for our U-Net model with bottleneck fusion for binary segmentation.}
    \label{tab:adaptions-study-semi-sup}
\end{table*}

\begin{table*}[t]
    \centering
    \begin{tabular}{l|p{1.7cm}  p{1.7cm} p{1.7cm} p{1.7cm}}
     & Dice & IoU & Precision & Recall \\
         \hline
         Our U-Net, no semi-supervision & 0.6781 & 0.5666 & 0.6945 & 0.6842\\
         Our U-Net, OWC-SSL & 0.6787 &  0.5672 &  \textbf{0.7146} &  0.7080 \\
         Our U-Net, OWC-SSL + uncertainty & \textbf{0.6884} & \textbf{0.5755} & 0.7139 & \textbf{0.7163}\\
    \end{tabular}
    \caption{Ablation study on semi-supervised learning for our U-Net model with bottleneck fusion for multiclass segmentation.}
    \label{tab:adaptions-study-semi-sup-multiclass}
\end{table*}

%%%%%%%%%%%%%%
\section{Discussion}\label{sec:discussion}
%%%%%%%%%%%%%%

In this work, we presented a semi-supervised deep learning-based segmentation approach designed to effectively integrate multi-modal data within a U-Net-based architecture for the segmentation of carotid artery vessel walls and plaques. With our experiments, we wanted to demonstrate the effectiveness of our approach and analyze the contribution of fusion strategies and semi-supervised learning to the segmentation performance in multi-sequence MRI data settings.
A key finding of our study is that the choice of fusion strategy has a significant effect on segmentation performance. In particular, our results demonstrate that bottleneck fusion outperforms early fusion in U-Net-based architectures, as it enables more effective integration of complementary information from different MRI sequences. An interesting observation is also that basic U-Net can achieve performance comparable to more advanced U-Net-based architectures simply by employing bottleneck fusion instead of early fusion. This finding underscores the importance of carefully selecting fusion points in deep learning architectures when working with multi-modal medical imaging data.

\begin{figure}
    \centering
    \begin{subfigure}{.45\linewidth}
    \includegraphics[width=\linewidth]{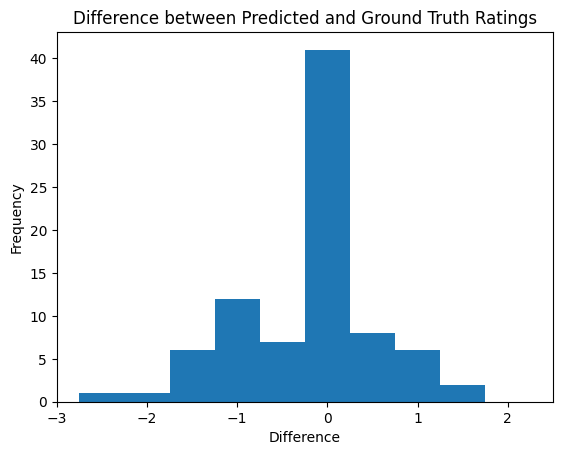}
    \caption{Binary}
    \end{subfigure}
    \hfill
    \begin{subfigure}{.45\linewidth}
    \includegraphics[width=\linewidth]{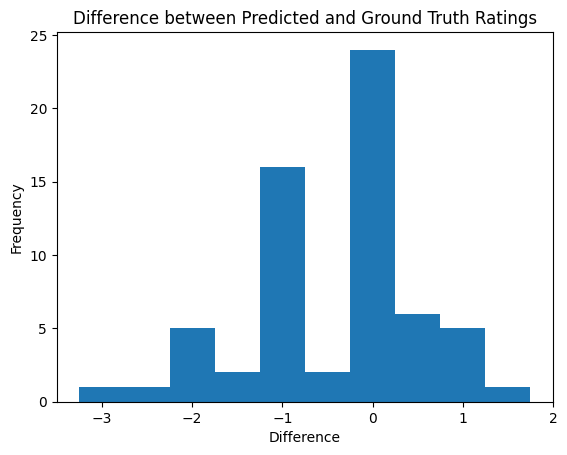}
    \caption{Multiclass}
    \end{subfigure}
    \caption{Histogram of average scores for predicted ratings minus ground truth ratings}
    \label{fig:hist}
\end{figure}

Our experiments have shown that our proposed model architecture outperformed basic U-Net for binary segmentation, especially for input but also for bottleneck fusion.
Notably, due to architectural enhancements such as SE-modules, deep supervision, residual blocks, and different up- and downsampling operations, our model exhibited greater robustness to the choice of fusion strategy compared to the standard U-Net. This suggests that our modifications help mitigate sensitivity to fusion points, making the model more adaptable to varying data characteristics.
For multiclass segmentation, we observed a slight decrease in segmentation performance compared to binary segmentation. 
However, this decline can be partially attributed to inconsistencies in the ground-truth annotations, where plaque and vessel wall boundaries were not always accurately delineated. In particular, the given ground truth masks, that were used for both loss computation and evaluation, contained some errors like missing mask pixels, inaccurate boundaries and wrong distinction between plaque and vessel wall. Non-perfect ground truth masks are a very common problem in medical image segmentation tasks and different experts or sites may have different methods for manually annotating anatomical structures, \citep{prbuge2005,atarce2023}. Manual annotations often suffer from the lack of reliability and reproducibility, and thus the true ground truth may need to be estimated from a collection of manual segmentations, \citep{wazowe2004}. Especially the annotation of pathological changes within vascular structures poses a very challenging task even for radiologists with several years of experience. Due to the error prone comparison masks, evaluation scores are relative, meaning that correctly identified pixels in the prediction masks may decrease the segmentation scores if they are wrong in the labels.
To assess the accuracy of the predicted segmentation masks from a different perspective, two expert radiologists conducted a blind evaluation. They were presented with randomly shuffled pairs of ground-truth and predicted masks from randomly selected patients and asked to determine which provided a better segmentation of the vessel wall and plaque—or if both were of equal quality. This approach aimed to provide an objective comparison of the predicted segmentation quality relative to expert-annotated ground-truth data. 84 binary and 63 multiclass predictions and ground truth segmentation masks were rated from 1 to 5 (where 5 was the best possible score and 1 the worst) by two expert radiologists, where one of the two was involved in the creation of ground truth annotations. The scores of the two raters were analyzed by averaging their rating scores per ROI-segmentation mask. For the inter-rater reliability, the Cohens Kappa coefficient was computed, leading to a score of $0.4406$ in the binary case and a score of $0.4915$ for the multiclass case. According to \citep{viga2005}, there is a moderate agreement between the two ratings, indicating reasonable consistency but also non-negligible variation, likely caused by subjective interpretation or the complexity of the segmentation task. This further stresses the necessity of a reliable segmentation method for these structures which remain a critical challenge in medial image analysis, particularly when dealing with multiple sequences. 
Concerning binary segmentation, the ratings from the predictions resulted in a mean score of $4.56$ with a standard deviation of $0.619$. For the ground truth masks, we achieved a mean score of $4.75$ with a standard deviation of $0.502$. The mean absolute difference of the ratings for the predicted and the ground truth masks was $0.500$. We further present a histogram of the difference (mean score of the predicted minus mean score of the ground truth ratings) in Figure~\ref{fig:hist}. Overall the scores of the predicted masks are slightly lower compared to those of the ground truth masks, however in most cases (images) the same score was achieved and in $16$ cases the predicted masks were perceived superior by the radiologists. 
The multiclass results yielded a mean score of $4.246$ with a standard deviation of $0.750$ for the predicted masks, and a mean score of $4.659$ with a standard deviation of $0.495$ for the ground truth masks, respectively. The mean absolute difference in the multiclass case was $0.714$. Concerning the multiclass histogram in Figure~\ref{fig:hist}, for $36$ predictions a better or equal score as in the ground truth masks was achieved. For $7$ predictions the ratings were lower by two points or more compared to the ground truth values. Reconsidering these lower-rated images, we found that in three of these cases the carotid artery was completely occluded (and thus not visible on the TOF sequence)—a condition which was not present elsewhere in our dataset. The others showed image (and movement) artifacts, which may have compromised the segmentation and differentiation between vessel wall and plaque. 
Keeping the fact in mind that evaluating multiclass predictions in a binary setting resulted in similar results as in the training of the binary case, the model is capable to segment plaque and vessel wall together but is less precise when separating the two tissues. A possible error case is, as mentioned before, the quality of ground truth segmentation masks. Over $6\%$ of the segmentation masks in the multiclass test set were rated with a score of $3.5$ or lower, presumably leading to a similar portion of less good ground truth masks in the training, negatively influencing the behavior of the model and the segmentation predictions. The low scores observed in the ground truth masks once again highlight the inherent subjectivity of manual segmentations and emphasize the need for objective, standardized approaches.

Despite the fact that semi-supervised learning only resulted in a small improvement in quantitative metrics like Dice and IoU, our approach still proves to be a valuable strategy for enhancing segmentation performance. The limited gain can be primarily attributed to the small number of available unlabeled patients. In particular, semi-supervised learning tends to show significant benefits when there is a large amount of unlabeled data available, ideally several times larger than the labeled dataset. This would allow the model to learn from a wider range of different structures and imaging conditions, ultimately enhancing its segmentation performance. In our case, the ratio between labeled and unlabeled data was almost the same. Moreover, due to the overall small dataset size, the additional unlabeled patients did not introduce much more variability, which may have further reduced the impact of semi-supervised learning.
Nevertheless, with a larger and more diverse dataset, semi-supervised learning is likely to yield greater improvements by leveraging additional unlabeled data. Further, the visual assessment of the results obtained by semi-supervised training have shown to be more robust when faced with diverse plaque morphologies and varying image quality compared to the ones obtained by fully supervised training. This means, especially in more complex cases, where stronger artifacts and less common plaque characteristics made segmentation more challenging, semi-supervised learning has shown to provide better segmentation results.
\\

\begin{figure}[p]
    \centering
    \begin{subfigure}{.9\linewidth}
    \includegraphics[width=\linewidth, trim = 0 50 0 20, clip]{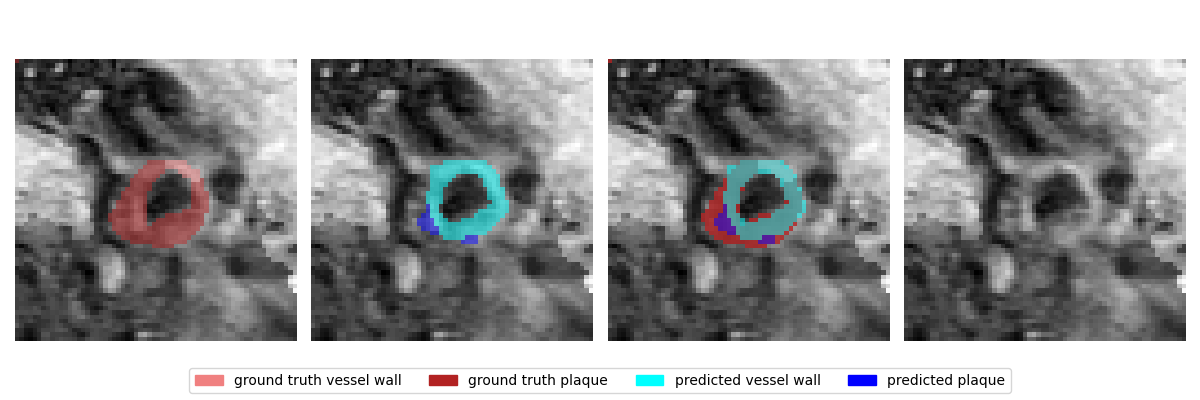} 
    \includegraphics[width=\linewidth, trim = 0 50 0 20, clip]{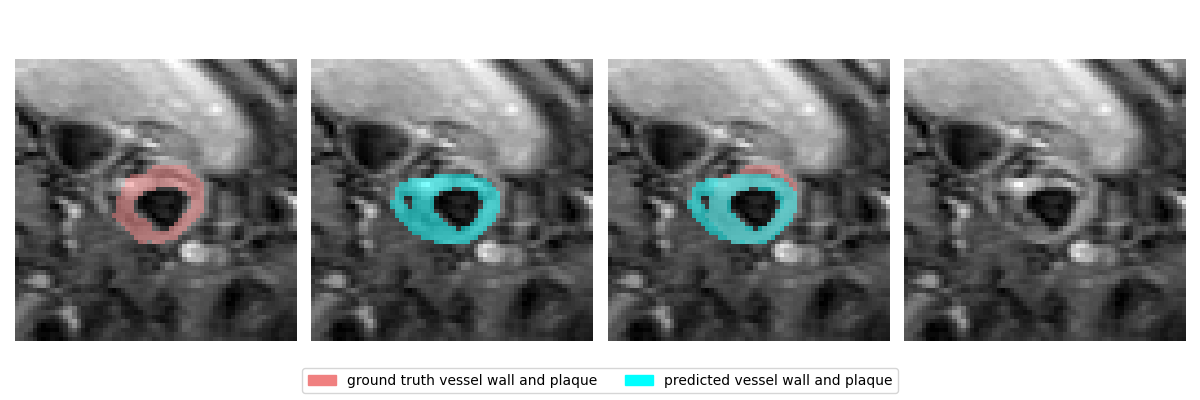}
    \caption{The predicted segmentation masks did not capture the whole plaque.} \label{not-full2}
    \end{subfigure}
    \begin{subfigure}{.9\linewidth}
    \includegraphics[width=\linewidth, trim = 0 50 0 20, clip]{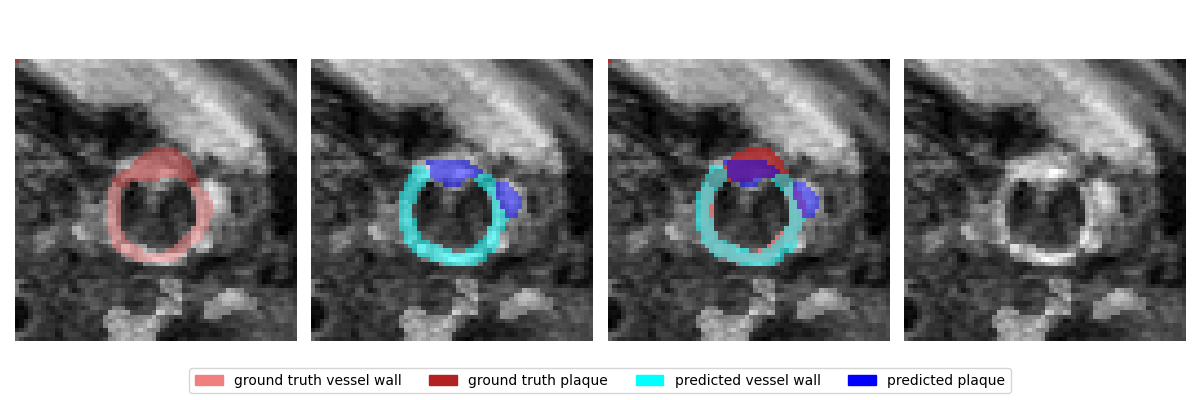}
    \includegraphics[width=\linewidth, trim = 0 50 0 20, clip]{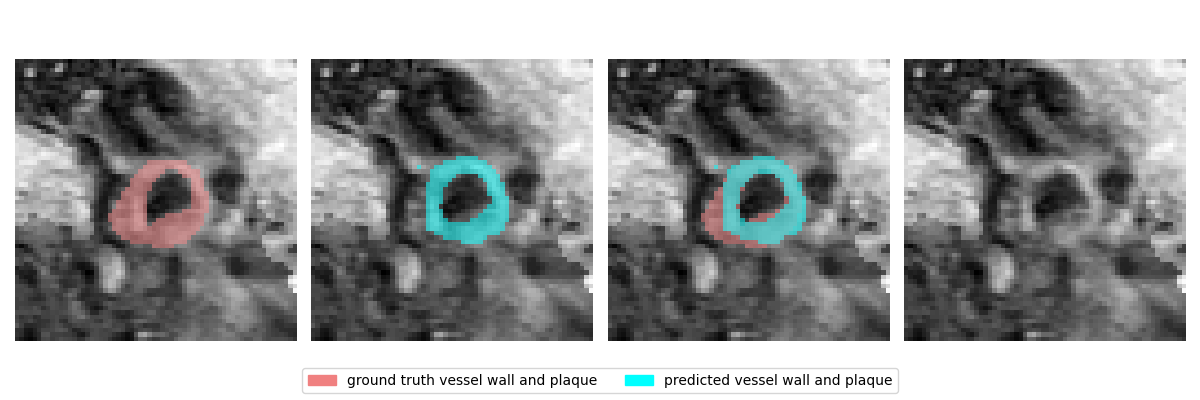}
    \caption{Examples of bad image quality slices.}
    \label{bad-quality2} 
    \end{subfigure}
    \begin{subfigure}{.9\linewidth}
    \includegraphics[width=\linewidth, trim = 0 50 0 20, clip]{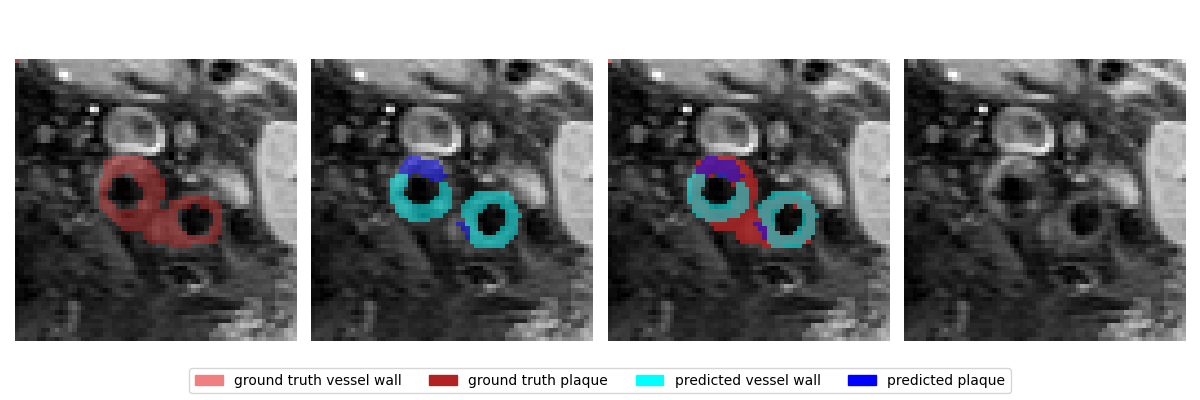}
    \caption{No distinction between vessel wall and plaque in ground truth mask.} \label{gt-only-plaque} 
    \end{subfigure}
    \caption{Example predictions from our U-Net model using bottleneck fusion and ground truth masks from expert radiologists for multiclass and binary segmentation. From left to right: Ground truth mask, predicted mask, their overlay, PDw image slice.}
    \label{fig:improvement}
\end{figure}

While the proposed approach yielded promising results, there still remains some room for improvements. One of the main limitations is the small dataset, consisting of MR-images from only 52 patients, with labeled data available only for 29 cases. Further, some MR-images were compromised by motion artifacts, such as those caused by swallowing, which obscures anatomical structures and thus diminishes segmentation accuracy (see Examples~\ref{bad-quality2}). 
We are aware that this limited dataset size poses significant challenges for training deep learning models efficiently. However, we deliberately chose this dataset due to its unique richness. The imaging data, including five distinct MRI sequences, is part of a dataset where also additional corresponding histopathological plaque classifications are available. This combination is exceedingly rare in clinical imaging studies and provides a valuable foundation for downstream radiomics analysis, enabling not only advanced tissue characterization but also the development of imaging biomarkers linked to histologically confirmed plaque vulnerability. As expert annotations showed considerable inter-observer variability, making consistent manual labeling difficult and limiting the reproducibility of radiomics-based assessments, we aimed to develop an automated segmentation approach that remains robust even in scenarios with limited data availability and complex anatomical structures.
Future efforts could focus on evaluating the proposed method using larger, higher-quality datasets. Additionally, improved ground-truth annotations could result in better segmentation masks and scores. As such, ground-truth masks should specifically differ between vessel wall and plaque (in Example~\ref{gt-only-plaque} the ground truth mask only contains plaque although there are also vessel wall regions), and include intra-vessel walls when the carotid artery bifurcated (as in the last example in Figure~\ref{fig:example_output}).

Further improvements could also be achieved by addressing registration mismatches in the MRI sequences. As suggested by \citep{wayu2024}, introducing small random translations and rotations across all sequences in the training step could help the model handle minor alignment errors in multi-modal data more effectively.
Another potential improvement could also be achieved by analyzing the fusion strategy employed in the decoder. In the current model, fusion maps are concatenated before being processed by a convolutional block. An alternative approach could involve weighting the feature maps with trainable parameters, allowing the model to learn the relative importance of different features dynamically. This could enhance the decoder’s ability to integrate multi-modal information more effectively and potentially improve segmentation performance.
Additionally, the proposed model could be extended to 3D with the additional difficulty that manual 2D slices often do not pose a smooth 3D object. Signed distance maps could be used here as proposed in \citep{wayu2024}. A signed distance map could also be beneficial for erroneous ground truth masks.

Post-processing of the predicted masks using conditional random fields or other methods that compare pixels inside or outside the boundary of the segmentation mask on similarity could further improve the results (see Examples~\ref{not-full2}). Also other properties such as connectivity of boundaries could be incorporated (see last example in Figure~\ref{fig:multiclass_unet_vs_marie}). 
Lastly, it would be interesting to investigate why bottleneck fusion yielded way better results than input fusion for the basic U-Net and what modification changed this behavior for our proposed U-Net.

%%%%%%%%%%%%%%
\section{Conclusion}\label{sec:conclusion}
%%%%%%%%%%%%%%

In this work, we presented a fully automated method for the segmentation of carotid artery vessel wall and plaques in multi-sequence MRI data using semi-supervised learning.
In both the localization and segmentation tasks, we introduced a novel multi-level multi-sequence version of U-Net architecture. Additionally, the localization model was enhanced by integrating prior knowledge on the number and positions of the carotid arteries. With our approach, we not only focused on designing an efficient U-Net structure but also on identifying the most effective fusion point for multi-modal data in U-Net-based architectures. 
Given the limited amount of labeled data and the complexity inherent in multi-sequence MRI data, we adopted a semi-supervised approach, extending the methodology of existing works on carotid artery segmentation and consistency regularization. Our approach leverages unlabeled data by enforcing the model to make consistent predictions under different geometric and photometric perturbations by employing a one-way consistency clean-teacher-perturbed-student framework. Further, we also included the idea of an uncertainty-aware scheme, ensuring that the student model learns only from meaningful and reliable predictions of the teacher model.
Comprehensive experiments demonstrated the effectiveness of our approach and showed that bottleneck fusion significantly improves segmentation performance in multi-sequence MRI data. Future work could explore larger datasets, higher-resolution images, and refined annotations to further enhance the accuracy of the proposed method.

\section*{Acknowledgments and Disclosure of Funding}
\noindent This work was supported by VASCage – Centre on Clinical Stroke Research. VASCage is a COMET Centre within the Competence Centers for Excellent Technologies (COMET) programme and funded by the Federal Ministry for Climate Action, Environment, Energy, Mobility, Innovation and Technology, the Federal Ministry of Labour and Economy, and the federal states of Tyrol, Salzburg and Vienna. COMET is managed by the Austrian Research Promotion Agency (Österreichische Forschungsförderungsgesellschaft).
\newline FFG Project number: 898252, Subproject: Imaging Biomarkers For Vascular Diseases And Vascular Aging.

% \section*{Data availability}
% \noindent Due to the sensitive and confidential nature of patient data used in this study, the dataset cannot be publicly shared. Access to the data is restricted in compliance with ethical and legal regulations.

\newpage

\vskip 0.2in
\bibliography{references}

\end{document}

%% file: makros.tex
\newcommand{\R}{{\mathbb{R}}}

\newcommand{\Tg}{T_\gamma^\mathrm{G}} % geometric transform
\newcommand{\Tp}{T_\phi^\mathrm{P}} % photometric transform
\newcommand{\Tgp}{T_\tau} % perturbation transform
\newcommand{\Tgpdef}{T_\gamma^\mathrm{G}\circ T_\phi^\mathrm{P}} % definition of perturbation transform
\newcommand{\paramG}{\gamma}
\newcommand{\paramP}{\phi}
\newcommand{\paramGP}{\tau}
\newcommand{\paramS}{\theta} % network parameter student model
\newcommand{\paramT}{{\theta'}} % network parameter teacher model
\newcommand{\paramSt}{\theta_t} % network parameter student model at timestep t
\newcommand{\paramTt}{{\theta_t'}} % network parameter teacher model at timestep t
\newcommand{\paramTtm}{{\theta_{t-1}'}} % network parameter teacher model at timestep t-1

\newcommand{\Ltotal}{\mathcal{L}^\mathrm{total}} % total loss
\newcommand{\Lsup}{\mathcal{L}^\mathrm{sup}} % supervised loss
\newcommand{\Lcon}{\mathcal{L}^\mathrm{con}} % consistency loss
\newcommand{\Lprior}{\mathcal{L}^\mathrm{prior}} % prior loss

\newcommand{\pred}{\mathrm{p}} % prediction
\newcommand{\gt}{\mathrm{y}} % ground truth

\newcommand{\img}{x} % image slice
\newcommand{\Xl}{X^\mathrm{L}} % concatenation of labeled images
\newcommand{\Xlu}{X} % concatenation of all images
\newcommand{\Y}{Y} % labels
\newcommand{\uncert}{\mathrm{u}} % uncertainty

\newcommand{\dimgh}{H} % dimension 1 of image
\newcommand{\dimgw}{W} % dimension 2 of image

\newcommand{\modelS}{\Phi_\paramS}

\newcommand{\modelT}{\Phi_\paramT}

%% file: main_manuscript.bbl
\begin{thebibliography}{47}
\providecommand{\natexlab}[1]{#1}
\providecommand{\url}[1]{\texttt{#1}}
\expandafter\ifx\csname urlstyle\endcsname\relax
  \providecommand{\doi}[1]{doi: #1}\else
  \providecommand{\doi}{doi: \begingroup \urlstyle{rm}\Url}\fi

\bibitem[Athanasiou et~al.(2023)Athanasiou, Arcos, and Cerquides]{atarce2023}
Georgios Athanasiou, Josep~Lluis Arcos, and Jesus Cerquides.
\newblock Enhancing medical image segmentation: Ground truth optimization through evaluating uncertainty in expert annotations.
\newblock \emph{Mathematics}, 11\penalty0 (17):\penalty0 3771, 2023.

\bibitem[Athiwaratkun et~al.(2019)Athiwaratkun, Finzi, Izmailov, and Wilson]{atfiiz2019}
Ben Athiwaratkun, Marc Finzi, Pavel Izmailov, and Andrew~Gordon Wilson.
\newblock There are many consistent explanations of unlabeled data: Why you should average, 2019.
\newblock URL \url{https://arxiv.org/abs/1806.05594}.

\bibitem[Ayg{\"{u}}n et~al.(2018)Ayg{\"{u}}n, Sahin, and {\"{U}}nal]{ayhu2018}
Mehmet Ayg{\"{u}}n, Yusuf~Huseyin Sahin, and G{\"{o}}zde~B. {\"{U}}nal.
\newblock Multi modal convolutional neural networks for brain tumor segmentation.
\newblock \emph{CoRR}, abs/1809.06191, 2018.
\newblock URL \url{http://arxiv.org/abs/1809.06191}.

\bibitem[Bai et~al.(2017)Bai, Oktay, Sinclair, Suzuki, Rajchl, Tarroni, Glocker, King, Matthews, and Rueckert]{baaksi2017}
Wenjia Bai, Ozan Oktay, Matthew Sinclair, Hideaki Suzuki, Martin Rajchl, Giacomo Tarroni, Ben Glocker, Andrew King, Paul~M. Matthews, and Daniel Rueckert.
\newblock Semi-supervised learning for network-based cardiac mr image segmentation.
\newblock In \emph{Medical Image Computing and Computer-Assisted Intervention - MICCAI 2017: 20th International Conference, Quebec City, QC, Canada, September 11-13, 2017, Proceedings, Part II}, page 253–260, Berlin, Heidelberg, 2017. Springer-Verlag.
\newblock ISBN 978-3-319-66184-1.
\newblock \doi{10.1007/978-3-319-66185-8_29}.
\newblock URL \url{https://doi.org/10.1007/978-3-319-66185-8_29}.

\bibitem[Buslaev et~al.(2020)Buslaev, Iglovikov, Khvedchenya, Parinov, Druzhinin, and Kalinin]{buigkh2020}
Alexander Buslaev, Vladimir~I Iglovikov, Eugene Khvedchenya, Alex Parinov, Mikhail Druzhinin, and Alexandr~A Kalinin.
\newblock Albumentations: fast and flexible image augmentations.
\newblock \emph{Information}, 11\penalty0 (2):\penalty0 125, 2020.

\bibitem[Chen et~al.(2020)Chen, Sun, Canton, Balu, Hippe, Zhao, Li, Hatsukami, Hwang, and Yuan]{chsu2020}
Li~Chen, Jie Sun, Gador Canton, Niranjan Balu, Daniel~S. Hippe, Xihai Zhao, Rui Li, Thomas~S. Hatsukami, Jenq-Neng Hwang, and Chun Yuan.
\newblock Automated artery localization and vessel wall segmentation using tracklet refinement and polar conversion.
\newblock \emph{IEEE Access}, 8:\penalty0 217603–217614, 2020.
\newblock ISSN 2169-3536.
\newblock \doi{10.1109/ACCESS.2020.3040616}.

\bibitem[Chen et~al.(2022)Chen, Wang, Zhang, Fung, Thai, Moore, Mannel, Liu, Zheng, and Qiu]{chwazh2022}
Xuxin Chen, Ximin Wang, Ke~Zhang, Kar-Ming Fung, Theresa~C Thai, Kathleen Moore, Robert~S Mannel, Hong Liu, Bin Zheng, and Yuchen Qiu.
\newblock Recent advances and clinical applications of deep learning in medical image analysis.
\newblock \emph{Medical image analysis}, 79:\penalty0 102444, 2022.

\bibitem[Chen et~al.(2023)Chen, Chen, Lin, Lin, Chen, Yang, Zhang, Li, Wang, and Huang]{chch2023stroke}
Ya-Fang Chen, Zhen-Jie Chen, You-Yu Lin, Zhi-Qiang Lin, Chun-Nuan Chen, Mei-Li Yang, Jin-Yin Zhang, Yuan-zhe Li, Yi~Wang, and Yin-Hui Huang.
\newblock Stroke risk study based on deep learning-based magnetic resonance imaging carotid plaque automatic segmentation algorithm.
\newblock \emph{Frontiers in Cardiovascular Medicine}, 10, 2023.

\bibitem[Cui et~al.(2019)Cui, Liu, Li, Guo, Li, Li, Wang, Zeng, and Ye]{culili2019}
Wenhui Cui, Yanling Liu, Yuxing Li, Meng-Hao Guo, Yiming Li, Xiuli Li, Tianle Wang, Xiangzhu Zeng, and Chuyang Ye.
\newblock Semi-supervised brain lesion segmentation with an adapted mean teacher model.
\newblock \emph{ArXiv}, abs/1903.01248, 2019.
\newblock URL \url{https://api.semanticscholar.org/CorpusID:67855384}.

\bibitem[Du et~al.(2024)Du, Zhuang, and Huang]{duzhhu2024}
Chengyang Du, Jie Zhuang, and Xinglu Huang.
\newblock Deep learning technology in vascular image segmentation and disease diagnosis.
\newblock \emph{Journal of Intelligent Medicine}, 2024.

\bibitem[Feng et~al.(2018)Feng, Nie, Wang, and Shen]{feniwa2018}
Zishun Feng, Dong Nie, Li~Wang, and Dinggang Shen.
\newblock Semi-supervised learning for pelvic mr image segmentation based on multi-task residual fully convolutional networks.
\newblock In \emph{2018 IEEE 15th International Symposium on Biomedical Imaging (ISBI 2018)}, pages 885--888, 2018.
\newblock \doi{10.1109/ISBI.2018.8363713}.

\bibitem[Girshick et~al.(2013)Girshick, Donahue, Darrell, and Malik]{gidoda2013}
Ross~B. Girshick, Jeff Donahue, Trevor Darrell, and Jitendra Malik.
\newblock Rich feature hierarchies for accurate object detection and semantic segmentation.
\newblock \emph{CoRR}, abs/1311.2524, 2013.
\newblock URL \url{http://arxiv.org/abs/1311.2524}.

\bibitem[Grandvalet and Bengio(2004)]{grbe2004}
Yves Grandvalet and Yoshua Bengio.
\newblock Semi-supervised learning by entropy minimization.
\newblock In \emph{Proceedings of the 18th International Conference on Neural Information Processing Systems}, NIPS'04, page 529–536, Cambridge, MA, USA, 2004. MIT Press.

\bibitem[Grubi{\v{s}}i{\'c} et~al.(2023)Grubi{\v{s}}i{\'c}, Or{\v{s}}i{\'c}, and {\v{S}}egvi{\'c}]{grorse2023}
Ivan Grubi{\v{s}}i{\'c}, Marin Or{\v{s}}i{\'c}, and Sini{\v{s}}a {\v{S}}egvi{\'c}.
\newblock Revisiting consistency for semi-supervised semantic segmentation.
\newblock \emph{Sensors}, 23\penalty0 (2):\penalty0 940, 2023.

\bibitem[Han et~al.(2024)Han, Sheng, Song, Liu, Qiu, Ma, and Liu]{hash2024}
Kai Han, Victor~S. Sheng, Yuqing Song, Yi~Liu, Chengjian Qiu, Siqi Ma, and Zhe Liu.
\newblock Deep semi-supervised learning for medical image segmentation: A review.
\newblock \emph{Expert Systems with Applications}, 245:\penalty0 123052, 2024.
\newblock ISSN 0957-4174.
\newblock \doi{https://doi.org/10.1016/j.eswa.2023.123052}.
\newblock URL \url{https://www.sciencedirect.com/science/article/pii/S0957417423035546}.

\bibitem[Howard et~al.(2017)Howard, Zhu, Chen, Kalenichenko, Wang, Weyand, Andreetto, and Adam]{hozhch2017}
Andrew~G. Howard, Menglong Zhu, Bo~Chen, Dmitry Kalenichenko, Weijun Wang, Tobias Weyand, Marco Andreetto, and Hartwig Adam.
\newblock Mobilenets: Efficient convolutional neural networks for mobile vision applications.
\newblock \emph{CoRR}, abs/1704.04861, 2017.
\newblock URL \url{http://arxiv.org/abs/1704.04861}.

\bibitem[Hu et~al.(2017)Hu, Shen, and Sun]{hushsu2017}
Jie Hu, Li~Shen, and Gang Sun.
\newblock Squeeze-and-excitation networks.
\newblock \emph{CoRR}, abs/1709.01507, 2017.
\newblock URL \url{http://arxiv.org/abs/1709.01507}.

\bibitem[Isensee et~al.(2021)Isensee, J\"ager, Kohl, Petersen, and Maier-Hein]{isja2021}
Fabian Isensee, Paul~F. J\"ager, Simon A.~A. Kohl, Jens Petersen, and Klaus Maier-Hein.
\newblock nnu-net: a self-configuring method for deep learning-based biomedical image segmentation.
\newblock \emph{Nature Methods}, 18:\penalty0 203 -- 211, 2021.
\newblock \doi{https://doi.org/10.1038/s41592-020-01008-z}.

\bibitem[Izmailov et~al.(2018)Izmailov, Podoprikhin, Garipov, Vetrov, and Wilson]{izpoga2018}
Pavel Izmailov, Dmitrii Podoprikhin, Timur Garipov, Dmitry Vetrov, and Andrew~Gordon Wilson.
\newblock Averaging weights leads to wider optima and better generalization.
\newblock \emph{arXiv preprint arXiv:1803.05407}, 2018.

\bibitem[Ker et~al.(2017)Ker, Wang, Rao, and Lim]{kewara2017}
Justin Ker, Lipo Wang, Jai Rao, and Tchoyoson Lim.
\newblock Deep learning applications in medical image analysis.
\newblock \emph{Ieee Access}, 6:\penalty0 9375--9389, 2017.

\bibitem[Khaniabadi et~al.(2023)Khaniabadi, Ibrahim, Huqqani, Khaniabadi, Sakim, and Teoh]{khibhu2023}
Shadi~Mahmoodi Khaniabadi, Haidi Ibrahim, Ilyas~Ahmad Huqqani, Farzad~Mahmoodi Khaniabadi, Harsa Amylia~Mat Sakim, and Soo~Siang Teoh.
\newblock Comparative review on traditional and deep learning methods for medical image segmentation.
\newblock In \emph{2023 IEEE 14th control and system graduate research colloquium (ICSGRC)}, pages 45--50. IEEE, 2023.

\bibitem[Khanna et~al.(2020)Khanna, Londhe, Gupta, and Semwal]{khlo2020}
Anita Khanna, Narendra~D. Londhe, S.~Gupta, and Ashish Semwal.
\newblock A deep residual u-net convolutional neural network for automated lung segmentation in computed tomography images.
\newblock \emph{Biocybernetics and Biomedical Engineering}, 40\penalty0 (3):\penalty0 1314--1327, 2020.
\newblock ISSN 0208-5216.
\newblock \doi{https://doi.org/10.1016/j.bbe.2020.07.007}.
\newblock URL \url{https://www.sciencedirect.com/science/article/pii/S0208521620300887}.

\bibitem[Li et~al.(2023)Li, Zheng, Zayed, Saffitz, Woodard, and Jha]{li2023carotid}
Ran Li, Jie Zheng, Mohamed~A Zayed, Jeffrey~E Saffitz, Pamela~K Woodard, and Abhinav~K Jha.
\newblock Carotid atherosclerotic plaque segmentation in multi-weighted mri using a two-stage neural network: advantages of training with high-resolution imaging and histology.
\newblock \emph{Frontiers in Cardiovascular Medicine}, 10:\penalty0 1127653, 2023.

\bibitem[Li et~al.(2020)Li, Yu, Chen, Fu, Xing, and Heng]{liyuch2020}
Xiaomeng Li, Lequan Yu, Hao Chen, Chi-Wing Fu, Lei Xing, and Pheng-Ann Heng.
\newblock Transformation-consistent self-ensembling model for semisupervised medical image segmentation.
\newblock \emph{IEEE transactions on neural networks and learning systems}, 32\penalty0 (2):\penalty0 523--534, 2020.

\bibitem[Mall et~al.(2023)Mall, Singh, Srivastav, Narayan, Paprzycki, Jaworska, and Ganzha]{masisr2023}
Pawan~Kumar Mall, Pradeep~Kumar Singh, Swapnita Srivastav, Vipul Narayan, Marcin Paprzycki, Tatiana Jaworska, and Maria Ganzha.
\newblock A comprehensive review of deep neural networks for medical image processing: Recent developments and future opportunities.
\newblock \emph{Healthcare Analytics}, page 100216, 2023.

\bibitem[Oktay et~al.(2018)Oktay, Schlemper, Folgoc, Lee, Heinrich, Misawa, Mori, McDonagh, Hammerla, Kainz, Glocker, and Rueckert]{oksc2018}
Ozan Oktay, Jo~Schlemper, Lo{\"{\i}}c~Le Folgoc, Matthew C.~H. Lee, Mattias~P. Heinrich, Kazunari Misawa, Kensaku Mori, Steven~G. McDonagh, Nils~Y. Hammerla, Bernhard Kainz, Ben Glocker, and Daniel Rueckert.
\newblock Attention u-net: Learning where to look for the pancreas.
\newblock \emph{CoRR}, abs/1804.03999, 2018.
\newblock URL \url{http://arxiv.org/abs/1804.03999}.

\bibitem[Prastawa et~al.(2005)Prastawa, Bullitt, and Gerig]{prbuge2005}
Marcel Prastawa, Elizabeth Bullitt, and Guido Gerig.
\newblock Synthetic ground truth for validation of brain tumor mri segmentation.
\newblock In \emph{International Conference on Medical Image Computing and Computer-Assisted Intervention}, pages 26--33. Springer, 2005.

\bibitem[Rahdar et~al.(2023)Rahdar, Ahmadi, Naseripour, Akhtari, Sedaghat, Hosseinabadi, Yarmohamadi, Hajihasani, and Mirshahi]{raahna20223}
Amir Rahdar, Mohamad~Javad Ahmadi, Masood Naseripour, Abtin Akhtari, Ahad Sedaghat, Vahid~Zare Hosseinabadi, Parsa Yarmohamadi, Samin Hajihasani, and Reza Mirshahi.
\newblock Semi-supervised segmentation of retinoblastoma tumors in fundus images.
\newblock \emph{Research Square}, 2023.
\newblock \doi{10.21203/rs.3.rs-2648324/v1}.
\newblock URL \url{https://doi.org/10.21203/rs.3.rs-2648324/v1}.

\bibitem[Redmon et~al.(2015)Redmon, Divvala, Girshick, and Farhadi]{redigi2015}
Joseph Redmon, Santosh~Kumar Divvala, Ross~B. Girshick, and Ali Farhadi.
\newblock You only look once: Unified, real-time object detection.
\newblock \emph{CoRR}, abs/1506.02640, 2015.
\newblock URL \url{http://arxiv.org/abs/1506.02640}.

\bibitem[Ronneberger et~al.(2015)Ronneberger, Fischer, and Brox]{rofibr2015}
Olaf Ronneberger, Philipp Fischer, and Thomas Brox.
\newblock U-net: Convolutional networks for biomedical image segmentation.
\newblock \emph{CoRR}, abs/1505.04597, 2015.
\newblock URL \url{http://arxiv.org/abs/1505.04597}.

\bibitem[Sharma and Aggarwal(2010)]{shag2010}
Neeraj Sharma and Lalit~Mohan Aggarwal.
\newblock Automated medical image segmentation techniques.
\newblock \emph{Journal of Medical Physics / Association of Medical Physicists of India}, 35:\penalty0 3 -- 14, 2010.
\newblock URL \url{https://api.semanticscholar.org/CorpusID:30824724}.

\bibitem[Tarvainen and Valpola(2017)]{tava2017}
Antti Tarvainen and Harri Valpola.
\newblock Mean teachers are better role models: Weight-averaged consistency targets improve semi-supervised deep learning results.
\newblock \emph{Advances in neural information processing systems}, 30, 2017.

\bibitem[Tsakanikas et~al.(2020)Tsakanikas, Siogkas, Mantzaris, Potsika, Kigka, Exarchos, Koncar, Jovanovic, Vujcic, Ducic, Pelisek, and Fotiadis]{tssi2020}
Vassilis~D. Tsakanikas, Panagiotis~K. Siogkas, Michalis~D. Mantzaris, Vassiliki~T. Potsika, Vassiliki~I. Kigka, Themis~P. Exarchos, Igor~B. Koncar, Marija Jovanovic, Aleksandra Vujcic, Stefan Ducic, Jaroslav Pelisek, and Dimitrios~I. Fotiadis.
\newblock A deep learning oriented method for automated 3d reconstruction of carotid arterial trees from mr imaging.
\newblock In \emph{2020 42nd Annual International Conference of the IEEE Engineering in Medicine and Biology Society (EMBC)}, page 2408–2411. IEEE, July 2020.
\newblock ISBN 978-1-7281-1990-8.
\newblock \doi{10.1109/EMBC44109.2020.9176532}.
\newblock URL \url{https://ieeexplore.ieee.org/document/9176532/}.

\bibitem[Viera and Garrett(2005)]{viga2005}
Anthony Viera and Joanne Garrett.
\newblock Understanding interobserver agreement: The kappa statistic.
\newblock \emph{Family medicine}, 37:\penalty0 360--3, 06 2005.

\bibitem[Wang et~al.(2024)Wang, Yu, Zhang, Lu, and Qian]{wayu2024}
Jian Wang, Fan Yu, Mengze Zhang, Jie Lu, and Zhen Qian.
\newblock A 3d framework for segmentation of carotid artery vessel wall and identification of plaque compositions in multi-sequence mr images.
\newblock \emph{Computerized Medical Imaging and Graphics}, 116:\penalty0 102402, September 2024.
\newblock ISSN 08956111.
\newblock \doi{10.1016/j.compmedimag.2024.102402}.

\bibitem[Wang et~al.(2017)Wang, Chen, Yuan, Liu, Huang, Hou, and Cottrell]{wach2017}
Panqu Wang, Pengfei Chen, Ye~Yuan, Ding Liu, Zehua Huang, Xiaodi Hou, and Garrison~W. Cottrell.
\newblock Understanding convolution for semantic segmentation.
\newblock \emph{CoRR}, abs/1702.08502, 2017.
\newblock URL \url{http://arxiv.org/abs/1702.08502}.

\bibitem[Wang and Yao(2023)]{waya2023}
Yu~Wang and Yudong Yao.
\newblock Application of artificial intelligence methods in carotid artery segmentation: a review.
\newblock \emph{IEEE Access}, 2023.

\bibitem[Warfield et~al.(2004)Warfield, Zou, and Wells]{wazowe2004}
Simon~K Warfield, Kelly~H Zou, and William~M Wells.
\newblock Simultaneous truth and performance level estimation (staple): an algorithm for the validation of image segmentation.
\newblock \emph{IEEE transactions on medical imaging}, 23\penalty0 (7):\penalty0 903--921, 2004.

\bibitem[Wu et~al.(2019)Wu, Xin, Yang, Sun, Xu, Zheng, and Yuan]{wuxi2019}
Jiayi Wu, Jingmin Xin, Xiaofeng Yang, Jie Sun, Dongxiang Xu, Nanning Zheng, and Chun Yuan.
\newblock Deep morphology aided diagnosis network for segmentation of carotid artery vessel wall and diagnosis of carotid atherosclerosis on black-blood vessel wall mri.
\newblock \emph{Medical Physics}, 46\penalty0 (12):\penalty0 5544–5561, December 2019.
\newblock ISSN 2473-4209.
\newblock \doi{10.1002/mp.13739}.

\bibitem[Xia et~al.(2024)Xia, Zheng, Zou, Luo, Tang, Li, and Jiang]{xizhzo2024}
Qingling Xia, Hong Zheng, Haonan Zou, Dinghao Luo, Hongan Tang, Lingxiao Li, and Bin Jiang.
\newblock A comprehensive review of deep learning for medical image segmentation.
\newblock \emph{Neurocomputing}, page 128740, 2024.

\bibitem[Xu and Zhu(2022)]{weqi2022}
Wenjing Xu and Qing Zhu.
\newblock A semantic segmentation method with emphasis on the edges for automatic vessel wall analysis.
\newblock \emph{Applied Sciences}, 12\penalty0 (14), 2022.
\newblock ISSN 2076-3417.
\newblock \doi{10.3390/app12147012}.
\newblock URL \url{https://www.mdpi.com/2076-3417/12/14/7012}.

\bibitem[Xu et~al.(2022)Xu, Yang, Li, Jiang, Jia, Gong, Mao, Zhang, Teng, Zhu, He, Wan, Liang, Li, Hu, Zheng, Liu, and Zhang]{wexi2022}
Wenjing Xu, Xiong Yang, Yikang Li, Guihua Jiang, Sen Jia, Zhenhuan Gong, Yufei Mao, Shuheng Zhang, Yanqun Teng, Jiayu Zhu, Qiang He, Liwen Wan, Dong Liang, Ye~Li, Zhanli Hu, Hairong Zheng, Xin Liu, and Na~Zhang.
\newblock Deep learning-based automated detection of arterial vessel wall and plaque on magnetic resonance vessel wall images.
\newblock \emph{Frontiers in Neuroscience}, 16, June 2022.
\newblock ISSN 1662-453X.
\newblock \doi{10.3389/fnins.2022.888814}.
\newblock URL \url{https://www.frontiersin.org/journals/neuroscience/articles/10.3389/fnins.2022.888814/full}.

\bibitem[Yeung et~al.(2022)Yeung, Sala, Schönlieb, and Rundo]{yesasc2022}
Michael Yeung, Evis Sala, Carola-Bibiane Schönlieb, and Leonardo Rundo.
\newblock Unified focal loss: Generalising dice and cross entropy-based losses to handle class imbalanced medical image segmentation.
\newblock \emph{Computerized Medical Imaging and Graphics}, 95:\penalty0 102026, 2022.
\newblock ISSN 0895-6111.
\newblock \doi{https://doi.org/10.1016/j.compmedimag.2021.102026}.
\newblock URL \url{https://www.sciencedirect.com/science/article/pii/S0895611121001750}.

\bibitem[Yu et~al.(2019)Yu, Wang, Li, Fu, and Heng]{yuwali2019}
Lequan Yu, Shujun Wang, Xiaomeng Li, Chi-Wing Fu, and Pheng-Ann Heng.
\newblock Uncertainty-aware self-ensembling model for semi-supervised 3d left atrium segmentation.
\newblock In \emph{Medical image computing and computer assisted intervention--MICCAI 2019: 22nd international conference, Shenzhen, China, October 13--17, 2019, proceedings, part II 22}, pages 605--613. Springer, 2019.

\bibitem[Zhang et~al.(2017)Zhang, Liu, and Wang]{zhli2017}
Zhengxin Zhang, Qingjie Liu, and Yunhong Wang.
\newblock Road extraction by deep residual u-net.
\newblock \emph{CoRR}, abs/1711.10684, 2017.
\newblock URL \url{http://arxiv.org/abs/1711.10684}.

\bibitem[Zhu et~al.(2021)Zhu, Wang, Teng, Chen, Huang, Xia, Mao, and Bai]{zhwa2021}
Chenglu Zhu, Xiaoyan Wang, Zhongzhao Teng, Shengyong Chen, Xiaojie Huang, Ming Xia, Lizhao Mao, and Cong Bai.
\newblock Cascaded residual u-net for fully automatic segmentation of 3d carotid artery in high-resolution multi-contrast mr images.
\newblock \emph{Physics in Medicine and Biology}, 66\penalty0 (4):\penalty0 045033, February 2021.
\newblock ISSN 0031-9155, 1361-6560.
\newblock \doi{10.1088/1361-6560/abd4bb}.

\bibitem[Zhu et~al.(2022)Zhu, Wang, Chen, Teng, Bai, Huang, Xia, Shao, Gu, and Sun]{zhu2022complex}
Chenglu Zhu, Xiaoyan Wang, Shengyong Chen, Zhongzhao Teng, Cong Bai, Xiaojie Huang, Ming Xia, Zhanpeng Shao, Zheng Gu, and Peiliang Sun.
\newblock Complex carotid artery segmentation in multi-contrast mr sequences by improved optimal surface graph cuts based on flow line learning.
\newblock \emph{Medical \& Biological Engineering \& Computing}, 60\penalty0 (9):\penalty0 2693--2706, 2022.

\end{thebibliography}
